\documentstyle[epsfig,12pt,preprint,tighten,aps]{revtex}
\begin{document}

\draft

\title{\rightline{{\tt September 1998}}
\rightline{{\tt UM-P-98/44}}
\rightline{{\tt RCHEP-98/10}}
\ \\
Implications of the
$\nu_\mu \to \nu_s$ solution to the atmospheric neutrino
anomaly for early Universe cosmology}
\author{R. Foot\footnote{Email: foot@physics.unimelb.edu.au}}
\address{School of Physics\\
Research Centre for High Energy Physics\\
The University of Melbourne\\
Parkville 3052 Australia\\
(foot@physics.unimelb.edu.au)}
\maketitle

\begin{abstract}
By numerically solving the quantum kinetic equations
we compute the range of parameters where the
$\nu_\mu \to \nu_s$ oscillation
solution to the atmospheric neutrino anomaly
is consistent with a stringent big bang nucleosynthesis 
(BBN) bound of $N_{eff}^{BBN} \stackrel{<}{\sim} 3.6$.
We show that this requires tau neutrino masses
in the range $m_{\nu_\tau} \stackrel
{>}{\sim} 4\ eV$ (for $|\delta m^2_{atm}| = 10^{-2.5}\ eV^2$).
We discuss the implications of this scenario for hot+cold dark
matter, BBN, and the anisotropy of the cosmic microwave 
background.

\end{abstract}
\newpage
\noindent
{\bf I Introduction}
\vskip 0.5cm
Neutrino physics continues to provide the most
promising window on physics beyond the standard model.
There are numerous indications for neutrino
oscillations including the atmospheric neutrino
anomaly, solar neutrino problem, LSND experiment,
structure formation in the early Universe, 
indications of $N_{eff}^{BBN} < 3$ etc.
For a recent review, see e.g. Ref.\cite{langacker}.

The atmospheric neutrino anomaly\cite{atm} has 
been confirmed by the superKamiokande experiment\cite{sk}.
The observed up-down asymmetries of the detected muons
indicate\cite{fvy1} that the simplest solution to 
this anomaly is either $\nu_\mu \to \nu_\tau$\cite{mut} or
$\nu_\mu \to \nu_s$\cite{mus,fv,model,ls,bp} oscillations
(although significant additional mixing with $\nu_e$
cannot currently be excluded if $\delta m^2
\stackrel{<}{\sim} 2 \times 10^{-3}\ eV^2$\cite{fvy2}).
In each case the oscillations are maximal or nearly
maximal ($0.80 \stackrel{<}{\sim} \sin^2 2\theta 
\stackrel{<}{\sim} 1$) and
\begin{equation}
4\times 10^{-4} (10^{-3}) \stackrel{<}{\sim} 
|\delta m^2_{atmos}|/eV^2 
\stackrel{<}{\sim} 10^{-2},
\label{a}
\end{equation}
for $\nu_\mu \to \nu_\tau$ ($\nu_\mu \to \nu_s$) 
oscillations\cite{fvy1}.
Although experimentally similar, the $\nu_\mu \to \nu_\tau$
and $\nu_\mu \to \nu_s$ oscillation solutions to the 
atmospheric neutrino anomaly can be experimentally 
distinguished\cite{dist}\footnote{There is 
an astrophysical/cosmological arguement which favours
$\nu_\mu \to \nu_s$ over $\nu_\mu \to \nu_\tau$. Ref.\cite{sciama} 
points out that the decaying neutrino theory\cite{sciama2} for 
the ionisation of hydrogen in the interstellar medium, in 
conjunction with the assumption that the cosmological constant 
is zero favours $\nu_\mu \to \nu_s$ over $\nu_\mu \to \nu_\tau$.}.  
These two solutions will
also have quite different implications for early
Universe cosmology.  Naively the $\nu_\mu \to \nu_s$ oscillation
solution to the atmospheric neutrino anomaly would
appear to lead to an effective number of four 
neutrinos\cite{naive} in the early Universe, which 
would make it difficult
to reconcile big bang nucleosynthesis (BBN) with the 
estimated primordial light element abundances (for a 
recent discussion, see Ref.\cite{ts}).  It should be stressed
that the ordinary-sterile neutrino oscillations
can only populate the sterile state provided that
the lepton number of the Universe is very small. 
Since the origin of the baryon and lepton asymmetries 
of the Universe are unknown
it is possible that a lepton asymmetry was created
at some early time $T \gg 100 \ MeV$.
Furthermore, if the lepton 
to photon ratio is larger than about $10^{-5}$
then the BBN bounds will be evaded\cite{fv1}.
Clearly this is one possibility.
However another possibility is that the
neutrino oscillations themselves generate the 
lepton number.  In fact, a careful study of the 
ordinary-sterile neutrino oscillations in the early Universe
reveal that the oscillations themselves typically generate 
a much larger asymmetry for a large range of 
parameters\cite{ftv,shi,fv2,fv3}\footnote{
It turns out that this asymmetry
is typically generated at temperatures
less than 100 MeV so that the quantitative calculations
of Ref.\cite{fv1} are not generally applicable to the
case where neutrino oscillations generate the
asymmetry.}.
As already discussed in 
Ref.\cite{fv2}, there is a concrete mechanism whereby the 
large neutrino oscillation generated asymmetry can prevent 
the (near) maximal $\nu_\mu \to \nu_s$ oscillations from 
populating the sterile neutrino in the early Universe.
This scenario requires the tau neutrino to be in the
$eV$ mass range and to mix slightly with the
sterile neutrino.  In this case the 
$\nu_\tau \to \nu_s$ oscillations generate a large 
tau neutrino asymmetry in the early Universe which suppresses 
$\nu_\mu \to \nu_s$ oscillations (the tau neutrino asymmetry 
can suppress $\nu_\mu \to \nu_s$ oscillations because the 
matter term for $\nu_\mu \to \nu_s$ oscillations also
contains a part which is proportional to the tau neutrino 
asymmetry of the background plasma).  The calculation is 
somewhat non-trivial because the $\nu_\mu \to \nu_s$ 
oscillations like to produce a large muon neutrino asymmetry 
which can compensate for the effect of the large tau
neutrino asymmetry in the matter term for $\nu_\mu \to \nu_s$
oscillations.  The computation of Ref.\cite{fv2} 
utilised an approximate solution to the
quantum kinetic equations (which was called 
the `static approximation' in Ref.\cite{fv2}) and
is valid provided that the system was 
smooth enough.  This is a very useful approximation
because it saves considerable CPU time as well
as giving more insight than the more complicated
quantum kinetic equations.  However as pointed out in 
Ref.\cite{fv2} this approximation is not always valid 
for the entire range of $\sin^2 2\theta_0, \ \delta m^2$ 
parameter space of 
interest.  In particular, if $\sin^2 2\theta_0$ is large 
enough then this approximation is generally not valid 
because the lepton asymmetry is created so rapidly.
Thus, the main purpose of this paper is to do a more
exact computation (by using the quantum kinetic
equations rather than the simpler static approximation) of 
the region of parameter space where the $\nu_\mu \to \nu_s$
solution to the atmospheric neutrino anomaly is
consistent with BBN (assuming $N^{BBN}_{eff} 
\stackrel{<}{\sim} 3.6$).
Also we will also include the entire range Eq.(\ref{a})
of $\delta m^2_{atmos}$.  If the sterile neutrinos are not 
populated by $\nu_\mu \to \nu_s$ oscillations then during the 
low temperature evolution of the neutrino asymmetries some
electron neutrino asymmetry will be transferred from the
tau neutrino asymmetry (due to $\nu_\tau \to \nu_e$ 
oscillations), which significantly affects BBN. This
observation was first made in Ref.\cite{fv3}
and we include a discussion of this effect here for 
completeness.  We will also discuss the 
implications for the hot+cold dark matter model
and the anisotropy of the cosmic microwave background (CMB).

For the purposes of this paper, we assume that there is
only one light sterile neutrino. Thus, we are considering a
four neutrino model.  Probably the simplest four neutrino 
model which can explain the atmospheric neutrino anomaly
through $\nu_\mu \to \nu_s$ oscillations,
and also explain the solar neutrino problem
is the model of Ref.\cite{ls}.
In this model the solar neutrino problem is explained by
the oscillations of $\nu_e \to \nu_\mu,\nu_s$
with parameters consistent
with the small angle MSW solution of the solar neutrino 
problem. Our results (as well as the previous 
work\cite{fv2,fv3}) is applicable to this model.
However this model does not seem to 
be compatible with the LSND anomaly\cite{lsnd}.
Actually, as we will discuss in more detail in
section VIII, our results indicate
that $N_{eff}^{BBN} < 4$ does not seem
to be consistent
with any four neutrino model which explains 
all three experimental anomalies.
Hence, if experimental data indicate that $\nu_\mu \to \nu_s$
oscillations are required to explain the atmospheric
neutrino data, and if the solar and LSND anomalies have
been correctly interpreted in terms of neutrino
oscillations, then $N_{eff}^{BBN} < 4$
actually suggests the need for more than four neutrinos.

Admittedly, the theory of neutrino oscillations in the 
early Universe
is quite a complicated subject. The readers who are
primarily interested in our results
can skip directly to Figures 2,3,4. These figures summarise
the main results of this paper.

The outline of this paper is as follows: In section II we
briefly review the phenomenon of neutrino oscillation 
generated neutrino asymmetry in the early Universe. In section
III we explicitly write down the quantum kinetic equations
which we will need latter on,
and we numerically solve them for some illustrative 
examples. In
section IV we obtain the region of parameter space
where the $\nu_\mu \to \nu_s$ oscillation solution of
the atmospheric neutrino anomaly is consistent with  a BBN
bound of $N_{eff}^{BBN} \stackrel{<}{\sim} 3.6$.
In section V, VI, VII we discuss the detailed implications 
of this
four neutrino scheme for BBN, hot+cold dark matter model,
and the anisotropy of the CMB.  In section VIII we conclude.

\vskip 0.5cm
\noindent
{\bf II Oscillation generated neutrino asymmetry in 
the early Universe}
\vskip 0.5cm
Our notation/convention for ordinary-sterile 
neutrino two state mixing is as follows, the weak
eigenstates ($\nu_{\alpha}, \nu_s$),
with $\alpha = e, \mu$ or $\tau$, are linear combinations
of two mass eigenstates ($\nu_a, \nu_b$):
\begin{equation}
\nu_{\alpha} = \cos\theta_0 \nu_a + \sin\theta_0 \nu_b,\
\nu_{s} = - \sin\theta_0 \nu_a + \cos\theta_0 \nu_b.
\end{equation}
Note that we define $\theta_0$ such that $\cos2 \theta_0 > 0$ and
we take the convention that $\delta m^2 \equiv m^2_b - m^2_a$.
The neutrino asymmetries are defined by, 
\begin{equation}
L_{\nu_\alpha} \equiv 
{n_{\nu_\alpha} - n_{\bar \nu_\alpha} \over n_\gamma},
\label{def}
\end{equation}
with $n_i$ being the number density of species $i$.
Finally, note that when we refer to a neutrino, sometimes we
will mean neutrinos and anti-neutrinos.
We hope the correct meaning will be clear from context.

It was shown in Ref.\cite{ftv} that for 
$\nu_\alpha \to \nu_s$ ($\alpha = e,\mu,$ or $\tau$)
oscillations (with $|\delta m^2|
\stackrel{>}{\sim} 10^{-4} \ eV^2$) the evolution of 
lepton number has the form (for small $L_{\nu_\alpha}$)
\begin{equation}
{dL_{\nu_\alpha}\over dt} \simeq C\left(L_{\nu_\alpha} + 
{\eta \over 2}\right),
\end{equation}
where $\eta$ is set by the relic nucleon number densities (and
is expected to be small, $\eta/2 \sim 10^{-10}$) and $C$ is a 
function of time, $t$ (or equivalently temperature, $T$). 
At high temperature $C$ is negative, so that 
$(L_{\nu_\alpha} + \eta/2) \simeq 0$ is an approximate 
fixed point.  However if $\delta m^2 < 0$, then $C$ changes 
sign at a particular temperature $T = T_c$. At this
temperature rapid exponential growth of 
neutrino asymmetry occurs (unless $\sin^2 2\theta_0$ is very
tiny, see Eq.(\ref{large}) below).
The temperature where $C$ changes sign was calculated to be\cite{ftv}
\begin{equation}
T_c \approx 16
\left({-\delta m^2\cos2\theta_0\over eV^2}\right)^{1\over 6}
\ MeV.
\label{dons}
\end{equation}
The generation of neutrino asymmetry occurs because 
the $\nu_\alpha \to \nu_s$ oscillation probability is
different to the $\bar \nu_\alpha \to \bar \nu_s$ oscillation
probability due to the matter effects in a CP asymmetric
background. As the asymmetry is created, the background
becomes more CP asymmetric because the neutrino asymmetries
contribute to the CP asymmetry of the background.

The phenomenon of neutrino oscillation generated neutrino
asymmetry was studied in more detail along with
some applications in Refs.\cite{shi,fv2,fv3,fv4,bfv,bvw}.
Ref.\cite{shi} calculated the region of parameter space
where the neutrino asymmetries were created under the
approximation that all the neutrinos have a common momentum
($p \sim 3.15T$). In Ref.\cite{fv2}, the neutrino momentum
spread was taken into account by developing an
approximate solution to the quantum kinetic equations.
Further studies of this approximation, along with
an alternative derivation appears in Ref.\cite{bvw}.
Ref.\cite{fv2} also contains a detailed study of
the temperature region
$T \sim T_c$ where the initial exponential
growth of the neutrino asymmetry occurs.
In Ref.\cite{fv3} the low temperature $T \stackrel{<}
{\sim} T_c$ evolution of the neutrino asymmetry was studied.
After the initial exponential growth of the neutrino asymmetry
occurs, the collisions and eventually MSW transitions combine
to keep the asymmetry growing.  This low temperature evolution 
of the asymmetry is approximately independent of 
$\sin^2 2\theta_0$ (assuming that $\sin^2 2\theta_0 \ll 1$). 
The `final' value of the asymmetry arises at
the temperature\cite{fv3} 
\begin{equation}
T^f_\nu \simeq 0.5(|\delta m^2|/eV^2)^{1/4}\ MeV.
\label{nau}
\end{equation}
The magnitude of the final value was calculated to be\cite{fv3},
\begin{eqnarray}
L_\nu^f/h &\simeq& 0.29\ for 
\ |\delta m^2|/eV^2 \stackrel{>}{\sim} 1000,
\nonumber \\
L_\nu^f/h &\simeq& 0.23\ for \ 3 \stackrel{<}{\sim} 
|\delta m^2|/eV^2 \stackrel{<}{\sim} 1000,\nonumber \\
L_\nu^f/h &\simeq& 0.35\ for \ 10^{-4} \stackrel{<}{\sim} 
|\delta m^2|/eV^2 \stackrel{<}{\sim} 3,
\label{go}
\end{eqnarray}
where $h \equiv (T_\nu^3/T_\gamma^3)$.  
Strictly these results are valid for $\nu_\alpha \to \nu_s$
oscillations in isolation (i.e. in the idealized case where there
are only the two flavour oscillations $\nu_\alpha \to \nu_s$
occurring). For the realistic case of
three ordinary and one sterile neutrino considered in this paper,
these results hold approximately for the $\nu_\tau \to \nu_s$
oscillations (assuming that 
$m^2_{\nu_\tau} \gg m^2_{\nu_\mu}, m^2_{\nu_e}$)\footnote{
The alternative case where $\nu_\mu$ and $\nu_\tau$ are 
approximately maximal mixtures of two mass eigenstates 
is discussed in Ref.\cite{bfv}.}.  This 
large neutrino asymmetry occurs for
a wide range of $\sin^2 2\theta_0, \
\delta m^2$\cite{fv2,shi},
\begin{equation}
5 \times 10^{-10}\left[ {eV^2 \over |\delta m^2|}\right]^{1 \over 6}
\stackrel{<}{\sim} \sin^2 2\theta_0 \stackrel{<}{\sim} 
4(2) \times 10^{-5}\left[{eV^2 \over |\delta m^2|}\right]^{1
\over 2}, \ |\delta m^2| \stackrel{>}{\sim} 10^{-4} eV^2,
\label{large}
\end{equation}
with $\delta m^2 < 0$ for $\nu_{\mu,\tau} \to \nu_s$ 
($\nu_{e} \to \nu_s$) oscillations.
Note that the upper bound on $\sin^2 2\theta_0$ 
in Eq.(\ref{large}) assumes that the energy density 
of sterile neutrinos, which arise in the period before 
the exponential growth of neutrino asymmetry occurs 
(i.e. $T \stackrel{>}{\sim} T_c$) is less than 
$0.6$ of a standard neutrino species. That is
$\delta N^{BBN}_{eff} \stackrel{<}{\sim} 0.6$ is
assumed in Eq.(\ref{large}). Because of the present
observational uncertainties in the primordial
light element abundances this bound is not
rigorous. It is nevertheless interesting to suppose 
that the bound is rigorous and to explore the consequences.  
In any case it is clear that the generation of large
neutrino asymmetries is quite a general phenomenon 
which occurs for a large region of parameter space
if light sterile neutrinos exist\footnote{
In the region of parameter space where 
$|\delta m^2| \ll 10^{-4}\ eV^2$, the evolution of 
lepton number is dominated by oscillations between
collisions and the lepton number tends to be oscillatory
\cite{ekm,ks,shi}. Note however that, if say
$\nu_\tau \to \nu_s$ oscillations have $|\delta m^2|
\stackrel{>}{\sim} 10^{-4}\ eV^2,$ then the final
value of $L_{\nu_{\tau}}$ is so large that 
oscillations of say $\nu_e \to \nu_s$ with
$|\delta m^2| \ll 10^{-4} \ eV^2$ are heavily suppressed
by the large matter effects (caused by the large
$L_{\nu_\tau}$) for $T \stackrel{>}{\sim}
0.5 \ MeV$, and thus such oscillations cannot
have any effect for BBN.}.

Oscillation generated large neutrino asymmetries can have 
several important implications for cosmology, including,
\vskip 0.5cm
\noindent
(1) The BBN bounds on ordinary - sterile neutrino
oscillations can be evaded\cite{fv2,fv4}.
\vskip 0.5cm
\noindent
(2) If electron asymmetry is generated, then
big bang nucleosynthesis can be directly 
affected though the modification of
nuclear reaction rates (such as 
$\nu_e + N \leftrightarrow P + e^-$, 
$\bar \nu_e + P \leftrightarrow N + e^+$)\cite{fv3,bfv}.
\vskip 0.5cm
\noindent
(3) The modification of the neutrino number densities
will have implications for structure formation and will 
affect the anisotropy of the cosmic microwave background.
\vskip 0.5cm

In this paper we will illustrate all three of these 
implications within the framework of the four neutrino
model which solves the atmospheric neutrino anomaly
by $\nu_\mu \to \nu_s$ oscillations.

\vskip 0.5cm
\noindent
{\bf III The quantum kinetic equations for 
ordinary - sterile neutrino 
oscillations in the early Universe}
\vskip 0.5cm
The density matrix\cite{merzb,bm,early} for an
ordinary neutrino, $\nu_\alpha$ 
($\alpha = e, \mu, \tau$), of momentum with magnitude $p$
oscillating with a sterile neutrino in the early Universe can
be parameterised as follows,
\begin{equation}
\rho_{\nu_\alpha} (p) = {1 \over 2} P_0(p)[I + {\bf P}(p).
{\bf \sigma}],\
\rho_{\bar \nu_\alpha }(p) = 
{1 \over 2} \bar P_0(p)[I + {\bf {\bar P}}(p).
{\bf \sigma}],
\end{equation}
where $I$ is the $2 \times 2$ identity matrix and
${\bf P}(p) = P_x(p){\bf \hat x} + P_y (p) {\bf \hat y}
+ P_z(p){\bf \hat z}$, ${\bf \sigma} = \sigma_x {\bf \hat x} + 
\sigma_y {\bf \hat y} + \sigma_z {\bf \hat z}$ (the 
$\sigma_i$ are the Pauli matrices).
It will be understood that the density matrices
and the quantities $P_i$ also depend on time $t$ or, equivalently, 
temperature $T$ (the time temperature relation 
for $m_e \stackrel{<}{\sim} T \stackrel{<}{\sim} m_\mu$ is
$dt/dT \simeq -M_P/5.5T^3$, where $M_P \simeq 1.22 \times 10^{22} \
MeV$ is the Planck mass).  

We will normalise the density matrix such that
the momentum distributions of $\nu_{\alpha}$, $\nu_s$ are given by
\begin{equation}
N_{\nu_{\alpha}} = {1 \over 2}P_0(p)[I + P_z(p)]N^{0}(p),
\ N_{\nu_{s}} = {1 \over 2}P_0(p)[I - P_z(p)]N^{0}(p),
\label{c}
\end{equation}
where
\begin{equation}
N^0(p) \equiv {1 \over 2\pi^2}{p^2 
\over 1 + exp\left({p \over T}\right) }.
\end{equation}
Note that there are analogous equations for the anti-neutrinos
(with ${\bf P}(p) \to {\bf \bar P}(p)$ and $P_0 \to \bar P_0$).
The evolution of $P_0(p)$ and ${\bf P}(p)$ [or $\bar P_0(p),\
{\bf \bar P}(p)$] are governed by the 
equations\cite{bm,bvw},
\begin{eqnarray}
{\partial{\bf P}(p) \over \partial t}& = &{\bf V}(p) \times 
{\bf P}(p) + [1 - P_z(p)]
[{\partial \over \partial t}ln P_0(p)]{\bf \hat z} 
\nonumber \\
&-& [D(p) + {\partial \over \partial t}
ln P_0(p)][P_x(p){\bf \hat x} + P_y(p){\bf \hat y}],
\label{b1}\\
{\partial P_0(p) \over \partial t}& \simeq &\Gamma (p)
\left[K(p) - {1\over 2}P_0(p)(1 + P_z(p))\right],
\label{b2}
\end{eqnarray}
where $D(p) = \Gamma (p)/2$ and $\Gamma (p)$ is
the total collision rate of the weak 
eigenstate neutrino of momentum $p$ with the background 
plasma\footnote{ From Ref.\cite{ekt,bvw} it is given by $\Gamma (p) = 
yG_F^2T^5(p/3.15T)$ where $y \simeq 4.0$ for $\nu = \nu_e$ and 
$y \simeq 2.9$ for $\nu = \nu_{\mu},\nu_{\tau}$ (for $m_e 
\stackrel{<}{\sim} T \stackrel{<}{\sim} m_\mu$).} 
and $K(p) \equiv N^{eq}(p)/N^0(p)$, with $N^{eq}(p)$ being 
the expected number of neutrinos in thermal equilibrium, i.e.  
\begin{equation}
N^{eq}(p) \equiv {1 \over 2\pi^2}{p^2 \over 
1 + exp\left({p+ \mu_{\nu_\alpha} \over T}\right)}.
\end{equation}
For anti-neutrinos, $\mu_{\nu_\alpha} \to \mu_{\bar \nu_{\alpha}}$ 
in the above equation.  The chemical potentials $\mu_{\nu_\alpha},
\ \mu_{\bar \nu_{\alpha}}$, depend on the lepton asymmetry.  
In general, for a distribution in thermal equilibrium
\begin{equation}
L_{\nu_{\alpha}} = {1 \over 4\zeta (3)}\int^{\infty}_0
{x^2 dx \over 1 + e^{x+\stackrel{\sim}{\mu_{\alpha}}}} - 
{1 \over 4\zeta (3)}\int^{\infty}_0
{x^2 dx \over 1 + e^{x+\stackrel{\sim}{\mu_{\bar \alpha}}}},  
\end{equation}
where $\stackrel{\sim}{\mu_\alpha} \equiv \mu_{\nu_\alpha}/T$
and $\stackrel{\sim}{\mu_{\bar \alpha}} \equiv 
\mu_{\bar \nu_{\alpha}}/T$. Expanding out the above equation,
\begin{equation}
L_{\nu_{\alpha}} \simeq -{1 \over 24\zeta (3)}\left[
\pi^2 (\stackrel{\sim}{\mu}_{\alpha} - 
\stackrel{\sim}{\mu}_{\bar \alpha})
-6(\stackrel{\sim}{\mu}_{\alpha}^2 - 
\stackrel{\sim}{\mu}_{\bar \alpha}^2)ln2
+ (\stackrel{\sim}{\mu}_{\alpha}^3 - 
\stackrel{\sim}{\mu}_{\bar \alpha}^3) \right],
\label{j1}
\end{equation}
which is an exact equation for $\stackrel{\sim}{\mu}_{\alpha} =
- \stackrel{\sim}{\mu}_{\bar \alpha}$, otherwise it holds to a good
approximation provided that $\stackrel{\sim}{\mu}_{\alpha,\bar \alpha}
\stackrel{<}{\sim} 1$. For $T \stackrel{>}{\sim} T^{\alpha}_{dec}$
(where $T^e_{dec} \approx 2.5$ MeV and $T^{\mu,\tau}_{dec} \approx 3.5
$ MeV are the chemical decoupling temperatures),
$\mu_{\nu_{\alpha}} \simeq - \mu_{\bar \nu_{\alpha}}$
because processes such as $\nu_{\alpha} + \bar \nu_{\alpha}
\leftrightarrow e^+ + e^-$ are rapid enough to make
$\stackrel{\sim}{\mu}_{\alpha} +\stackrel{\sim}{\mu}_{\bar \alpha} 
\ \simeq \ \stackrel{\sim}{\mu}_{e^+} + \stackrel{\sim}{\mu}_{e^-} 
\simeq 0$.
However, for $1 MeV \stackrel{<}{\sim} T \stackrel{<}{\sim}
T^{\alpha}_{dec}$, weak interactions are rapid enough to 
approximately thermalise the neutrino momentum distributions,
but not rapid enough to keep the neutrinos in chemical equilibrium
\footnote{The chemical and thermal decoupling temperatures are so
different because the inelastic collision rates are much
less than the elastic collision rates. See e.g. Ref.\cite{ekt} for
a list of the collision rates.}.
In this case, the value of $\stackrel{\sim}{\mu}_{\alpha}$ is
approximately frozen at $T \simeq T^{\alpha}_{dec}$ (taking
for definiteness $L_{\nu_\alpha} > 0$), while the anti-neutrino 
chemical potential $\stackrel{\sim}{\mu}_{\bar \alpha}$
continues increasing until $T \simeq 1$ MeV.

The form of the evolution equation, Eq.(\ref{b1}) is very easy 
to understand.  The ${\bf V} \times {\bf P}$ term 
is simply the quantum mechanical coherent evolution of 
the states (this is derived in many
text books, see e.g. Ref.\cite{merzb}). The damping term
($D[P_x {\bf \hat x} + P_y {\bf \hat y}$]) is just the destruction
of the coherence of the ensemble due to collisions. The only suprising
thing is that $D(p)$ is {\it half} of the collision
frequency i.e. $D(p) = \Gamma (p)/2$ [rather than 
$\Gamma (p)$]\cite{stod}.  The rate of change of $P_0$ is 
due {\it only} to the repopulation 
of ordinary neutrinos due to weak interactions (since, from Eq.(\ref{c}),
$P_0(p) = (N_{\nu_\alpha} + N_{\nu_s})/N^0$, which is obviously
unchanged by the $\nu_\alpha \to \nu_s$ oscillations). 
Note that Eq.(\ref{b2}) is approximate and holds provided that
the distributions of the neutrinos 
and background particles are in near thermal equilibrium
(see Ref.\cite{bvw} for a detailed derivation of 
this formula).  Finally the $\partial lnP_0/\partial t$ 
terms arise because the repopulation does
not populate sterile states or mixtures of states.
[For example, because the repopulation does not affect the number
of sterile states, it follows that ${\partial P_z \over \partial t}
|_{repopulation} = (1-P_z){\partial lnP_0 \over \partial t}$, 
since ${1\over 2}P_0(1-P_z)$ 
is unchanged by the re-population].

The quantity ${\bf V}(p)$ is given by\cite{bm,early}
\begin{equation} 
{\bf V}(p) = \beta (p) {\bf \hat x} + \lambda (p) {\bf \hat z},
\end{equation}
where $\beta (p)$ and $\lambda (p)$ are
\begin{equation}
\beta (p) = {\delta m^2 \over 2p}\sin 2\theta_0,
\ \lambda (p) = -{\delta m^2 \over 2p}[\cos2\theta_0 - b(p)
\pm a(p)],
\label{sf}
\end{equation}
in which the $+(-)$ sign corresponds to neutrino (anti-neutrino)
oscillations. The dimensionless variables $a(p)$ and $b(p)$ 
contain the matter effects\cite{msw} (more precisely they are the
matter potential divided by $\delta m^2/2p$).
For $\nu_{\alpha} \to \nu_s$ oscillations 
$a(p), b(p)$ are given by\cite{rn} 
\begin{equation}
a(p) \equiv {-4\zeta(3)\sqrt{2}G_FT^3L^{(\alpha)}p
\over \pi^2\delta m^2},
\ b(p) \equiv {-4\zeta (3) \sqrt{2} G_F T^4 A_{\alpha} p^2 \over
\pi^2 \delta m^2 M_W^2},
\label{sal}
\end{equation}
where $\zeta (3) \simeq 1.202$ is the Riemann zeta function
of 3, $G_F$ is the Fermi constant, $M_W$ is the 
$W-$boson mass, $A_e \simeq 17$ and $A_{\mu, \tau} \simeq 4.9$ 
(for $m_e \stackrel{<}{\sim} T \stackrel{<}{\sim} m_\mu$).
The quantity $L^{(\alpha)}$ is given by
\begin{equation}
L^{(\alpha)} = L_{\nu_\alpha} + L_{\nu_e} + L_{\nu_\mu}
+ L_{\nu_\tau} + \eta,
\end{equation}
where $\eta$ is a small term due to the asymmetry of the 
electrons and nucleons and is expected
to be very small, $\eta \sim 5\times 10^{-10}$.
Recall that the neutrino asymmetry is defined in Eq.(\ref{def}).
For example, the number density of $\nu_\alpha$ is
\begin{equation}
n_{\nu_\alpha} = \int^{\infty}_{0} N_{\nu_\alpha}dp =
\int^{\infty}_0 {1 \over 2} P_0 (1 + P_z) N^0 dp.
\end{equation}
Note that in the interests of notational simplicity,
in the above equation and in the following discussion, 
the functional dependence of $P_i, N^0, \beta, \lambda, D$ 
on the neutrino momentum will not always be made explicit.
The rate of change of lepton number is given by
\begin{equation}
{dL_{\nu_\alpha} \over dt} =
{d \over dt}\left[ {n_{\nu_\alpha} - n_{\bar \nu_\alpha}\over
n_\gamma}\right]
= -{d \over dt}\left[ {n_{\nu_s} - n_{\bar \nu_s}\over
n_\gamma}\right].
\end{equation}
Thus, using Eq.(\ref{c}),
\begin{equation}
{dL_{\nu_\alpha} \over dt} =
{-1 \over 2n_\gamma}\int 
\left[{\partial P_0 \over \partial t}(1 - P_z) - 
P_0 {\partial P_z \over \partial t} 
- {\partial \bar P_0 \over \partial t}(1 - \bar P_z) + 
\bar P_0 {\partial\bar P_z \over \partial t} \right]N^0dp.
\label{wed}
\end{equation}
It is possible to numerically integrate the
system of coupled differential equations,
Eq.(\ref{b1},\ref{b2},\ref{wed}), although it 
is (CPU) time consuming. Unfortunately we do not have 
a parallel supercomputer handy. We therefore employ the
useful time saving approximation 
of integrating the oscillation equations, Eq.(\ref{b1}), 
in the region around the MSW resonance (we will
define this region precisely in a moment).
Actually away from the resonance the oscillations
are typically suppressed by the matter effects or by
$\sin^2 2\theta_0$ (or both).  Thus, this should 
be a good approximation, which we can check by taking larger 
slices of momentum space around the resonance.

In the adiabatic approximation, the oscillation probability 
for $\nu_\alpha \to \nu_s$ is given by the formula,
\begin{equation}
{\cal P} = \sin^2 2\theta_m \langle \sin^2 {\tau \over 2L_m}
\rangle.
\label{uu}
\end{equation}
Of course this formula is only valid provided that $\sin2\theta_m$
is approximately constant on the interaction time scale ($1/D$).
In Eq.(\ref{uu}), the brackets $\langle ... \rangle$ denotes the 
average over the interaction times $\tau$, and $\sin^2 2\theta_m,
L_m$ are the matter mixing angle and oscillation length
respectively. In terms of $\beta, \lambda, D$,
\begin{eqnarray}
\sin^2 2\theta_m &=& {\beta^2 \over \beta^2 + \lambda^2},
\ L^2_m = {1 \over \beta^2 + \lambda^2},
\nonumber \\ 
\langle \sin^2 {\tau \over 2L_m}\rangle &=&
D\int^{\infty}_0 e^{-\tau D}
\sin^2 {\tau \over 2L_m} d\tau. 
\end{eqnarray}
It is straightforward to show that\cite{fv2}
\begin{equation}
{\cal P} = {1 \over 2}{\beta^2 \over
\beta^2 + D^2 + \lambda^2}. 
\label{dff}
\end{equation}
Thus the oscillation probability has the MSW resonance when
$\lambda (p) = 0$ (note that for notational
convenience we call this the `MSW' resonance even
if $D^2 \gg \beta^2$). Solving the equation, 
$\lambda (p=p_{res}) = 0$ we find,
\begin{equation}
p_{res} =
{X_2 \over 2X_1} + \sqrt{\left( {X_2 \over 2X_1}\right)^2 + {
\cos2\theta_0 \over X_1}},
\label{f}
\end{equation}
where $X_1 \equiv b(p)/p^2$ and $X_2 \equiv a(p)/p$
(note that $X_i$ are independent of $p$).
The resonance width in momentum space, $\Delta$, can be obtained
from $\lambda(p=p_{res}\pm {1\over 2}\Delta)^2 \approx 
\beta^2 + D^2$,
\begin{equation}
\Delta \simeq {4p_{res} \over |\delta m^2|}
{\sqrt{\beta^2 + D^2} \over (2p_{res}X_1 - X_2)},
\label{g}
\end{equation}
where $\beta, D$ are evaluated at $p=p_{res}$.
For anti-neutrinos, $X_2 \to -X_2$ in Eqs.(\ref{f},\ref{g}).

For the numerical work the continuous variable $p/T$ is 
replaced by a finite set of momenta $x_n \equiv p_n/T$
(with $n=1,2,...,N$) on a logarithmically spaced mesh.
The variables ${\bf P}(p)$ and $P_0(p)$
are replaced by the set of $N$ variables ${\bf P}(x_n)$ 
and $P_0(x_n)$. The evolution of each of these variables is
governed by Eqs.(\ref{b1},\ref{b2}), 
where for each value of $n$, the variables
${\bf V}(p)$ and $D(p)$ are replaced by ${\bf V}(x_n)$ and
$D(x_n)$. The anti-neutrinos are similarly treated.
Thus the oscillations of the neutrinos and anti-neutrinos
can be described by $8N$ simultaneous differential equations.
For each time step the lepton number is computed from Eq.(\ref{wed})
and the chemical potentials are obtained from Eq.(\ref{j1}).
In order to integrate first order differential equations initial
conditions must be specified.
At very high temperatures the oscillations are 
heavily suppressed by the collisions so we take
$P_{x,y}(p) = 0, \bar P_{x,y}(p) = 0$ initially
[to see this observe that at high temperature, 
$D \gg \beta$ so that ${\cal P} \to 0$ from 
Eq.(\ref{dff})]. Also, we assume that there is initially
a negligible number of sterile neutrinos which means that
we take $P_z(p) = \bar P_z(p) = 1$ and $P_0(p) = 
\bar P_0(p) = 1$ initially. The value of
our initial temperature must be significantly higher than
the temperature where the exponential growth of lepton number occurs
[we typically used $T_{initial} \sim 4 T_c$, where $T_c$ is given in
Eq.(\ref{dons})].  In our numerical work we integrate Eq.(\ref{b1})
around the region 
\begin{equation}
p_{res} - f{\Delta\over 2} \ < \ p \ < 
\ p_{res} + f{\Delta \over 2},
\label{stu}
\end{equation}
with $f = 7$.  Note that repopulation equation Eq.(\ref{b2})
needs to be integrated over the entire region in momentum space,
which we typically approximated to be $0.001 < p/T < 20$. We also
take into account the damping in the momentum 
region away from resonance.  That is for the momentum region, 
\begin{equation}
p \ < \ p_{res} - f{\Delta\over 2} \ \ and \ \  p \ > 
\ p_{res} + f{\Delta \over 2},
\end{equation}
we neglect the oscillations and Eqs.(\ref{b1},\ref{b2})
are truncated to, 
\begin{eqnarray}
{\partial P_{x,y}(p) \over \partial t}
&=& -\left(D(p) + {\partial \over \partial t}lnP_0(p)\right)
P_{x,y}(p), \nonumber \\
{\partial P_z(p) \over \partial t} &=& \left(
1 - P_z(p)\right){\partial \over \partial t}lnP_0(p),\nonumber \\ 
{\partial P_0(p) \over \partial t}& \simeq &\Gamma (p)
\left[K(p) - {1\over 2}P_0(p)(1 + P_z(p))\right],
\end{eqnarray}
and similarly for $\bar P_{x,y,z}(p), \ \bar P_0(p)$.
We checked the stability of our results by integrating
Eq.(\ref{b1}) over larger slices of
momentum space [i.e. taking $f > 7$].

We have numerically solved these equations
for three illustrative examples. 
In Figure 1 we have plotted $|L_{\nu_\tau}|/h$ versus $T$
(recall that $h \equiv T_{\nu}^3/T_{\gamma}^3$) 
for $\nu_\tau \to \nu_s$ oscillations. 
We have chosen $\delta m^2/eV^2 = -0.5,\ -50,\ -5000$
and $\sin^2 2\theta_0 = 10^{-8}$ with
initial condition $L_{\nu_\alpha} = 0$ at $T_{initial} = 300\ MeV$.
In these examples the rapid exponential growth of lepton number
occurs when $T_c\simeq 16, 37, 80\ MeV$. In the region $T > T_c$, 
$L^{(\alpha)} \to 0$ is an approximate fixed point and this
explains why $L_{\nu_{\alpha}} \to -\eta/2 \simeq -2.5 \times 10^{-10}$.
In fact this behaviour occurs provided that the initial
value of $L_{\nu_\alpha}$ is less \footnote{
For initial values of $L_{\nu_\alpha} \stackrel{>}{\sim}
10^{-5}$ the oscillations are not able to drive $L^{(\alpha)} \to 0$.   
In this case the oscillations do not significantly destroy the
initial asymmetry in the region $T > T_c$.
In the lower temperature region $T < T_c$ the oscillations
increase the size of the asymmetry and the final value of the
asymmetry is also given approximately by Eq.(\ref{go}) (assuming
of course that the magnitude of the initial asymmetry 
was less than $L^f_\nu$ to begin with).} than about $10^{-5}$.

The behaviour illustrated in Figure 1 is well understood\cite{fv2,fv3}
and quite general provided that $\delta m^2 < 0$, $|\delta m^2|
\stackrel{>}{\sim} 10^{-4} \ eV^2$ and $\sin^2 2\theta_0$
is within the wide range, Eq.(\ref{large}).
In the region before the exponential growth of lepton number
occurs ($T \stackrel{>}{\sim} T_c$), $L^{(\alpha)} \approx 0$
(which means that $X_2 \approx 0$) so that the MSW oscillation
resonance for neutrinos $p_{res}^{\nu}$ has approximately
the same value as the MSW oscillation resonance for anti-neutrinos 
($p_{res}^{\bar \nu}$). As the neutrino asymmetry is created (taking
$L_{\nu_\alpha} > 0$ for definiteness), $p_{res}^{\nu}/T \gg 1$ 
(and typically $ \stackrel{>}{\sim} 20$) and $p_{res}^{\bar \nu}/T 
< 1$ (and typically $\stackrel{<}{\sim} 0.2$)\footnote{
See Ref.\cite{fv3} for a figure illustrating the
evolution of $p^{\nu, \bar \nu}_{res}$ for an example.}.
In the region after the exponential growth ($T \stackrel{<}{\sim}
T_c$) the collisions keep $p_{res}^{\bar \nu}/T < 1$. For
lower temperatures $T \stackrel{<}{\sim} T_c/2$, the $b(p)$
term [in Eq.(\ref{sal})] can be approximately neglected 
(relative to the $a(p)$ term) and
\begin{equation}
{p_{res}^{\bar \nu}\over T} \simeq {1\over TX_2}  
\ \ \alpha \ \ {1 \over T^4L^{(\tau)}}.
\label{fkd}
\end{equation}
As $T$ increases, $p_{res}^{\bar \nu}/T$ increases and
MSW transitions convert $\bar \nu_{\alpha} \to \bar \nu_s$
at the momentum $p = p_{res}^{\bar \nu}$.
The effect of this is to keep $L_{\nu_\alpha}$ growing
until all of the $\bar \nu$ have passed through the
MSW resonance\footnote{Note that the MSW transitions
keep $p_{res}^{\bar \nu}/T$ approximately constant
from $T \sim T_c/2$ until $L_{\nu_\tau}$ is
quite large ($\sim 10^{-2}$). From Eq.(\ref{fkd}),
$p_{res}^{\bar \nu}/T \approx \ constant$,
implies that $L_{\nu_\tau}$ is proportional to $1/T^4$.
This explains why ${dlogL_{\nu_\tau} \over dlog T}
\simeq -4,$ in the region after the exponential growth
of lepton number occurs for the examples in Figure 1.}. 
If there were no re-population effects (which is 
approximately true for the case of low $|\delta m^2| 
\stackrel{<}{\sim} 3\ eV^2$, where the asymmetry doesn't 
become large until low temperatures, $T \stackrel{<}{\sim} 
1 \ MeV$) then the expected asymmetry is $L_{\nu_\alpha} 
\approx n_{\nu_\alpha}/n_{\gamma} \simeq 0.375$ (since all 
of the $\bar \nu_\alpha$ have been converted into $\bar 
\nu_s$).  However in the case where $L_{\nu_{\alpha}}$ 
becomes very large in the region above about 1 MeV, 
repopulation effects must be taken into account, and the 
final value of the asymmetry is typically reduced somewhat
[see Eq.(\ref{go})].  Thus, this explains why the final 
asymmetry is so large and approximately independent of 
$\sin^2 2\theta_0$ so long as it is small.
Note that in the case of relatively large $\sin^2 2\theta_0$,
significant number of $\nu_s, \bar \nu_s$ are populated
in the region $T \stackrel{>}{\sim} T_c$. In this case,
the MSW transitions at low temperature are less effective
at creating $L_{\nu_{\alpha}}$. Indeed, in the limit
where the $\nu_s, \bar \nu_s$ are fully populated,
MSW transitions would simply interchange equal numbers
of $\bar \nu_{\alpha}$ with $\bar \nu_s$ and thus there
would be no significant generation of neutrino asymmetry
in this case.

Note that for the examples in Figure 1, the sign of $L_{\nu_\tau}$
changed at the temperature $T\simeq T_c$.
This behaviour is expected (see Ref.\cite{fv2} for  
a discussion of this point).  Actually our numerical
integration of the quantum kinetic equations reveals 
that for very large $\sin^2 2\theta_0$, 
oscillations of $L_{\nu_\tau}$ occur. 
In this case, it would presumably be impossible
to predict the sign of $L_{\nu_\tau}$ (see also
the discussion in Ref.\cite{shi}).
Finally note that this result was not found in Ref.\cite{fv2}
because the static approximation 
developed there is not valid for $\sin^2 2\theta_0$ 
sufficiently large (due to the extremely rapid exponential 
growth of lepton number).

\vskip 1.5cm
\noindent
{\bf IV Consistency of $\nu_\mu \to \nu_s$ solution of the
atmospheric neutrino anomaly with BBN} 
\vskip 0.5cm
We now turn to the main issue of this paper, that is,
to accurately determine the region of parameter space where the
$\nu_\mu \to \nu_s$ oscillation solution to the atmospheric
neutrino anomaly is consistent with a stringent BBN bound
of $N^{BBN}_{eff} \stackrel{<}{\sim} 3.6$.

Let us begin by introducing some notation.
We will denote the parameters, $\sin^22\theta_0, \delta m^2
$ and the matter terms $a(p),b(p)$ [Eq.(\ref{sal})] 
for $\nu_\alpha \to \nu_s$ oscillations
by $\sin^2 2\theta_0^{\alpha s}, \delta m^2_{\alpha s}$
and $a_{\alpha s}(p), b_{\alpha s}(p)$ respectively 
($\alpha = e,\mu, \tau$). 
We are assuming that there is only one sterile neutrino
which mixes maximally or near maximally with the muon neutrino
so that the atmospheric neutrino anomaly is solved, i.e.
\begin{equation}
\sin^2 2\theta_0^{\mu s} \simeq 1,
\ 10^{-3} \stackrel{<}{\sim} 
|\delta m^2_{\mu s}|/eV^2 \stackrel{<}{\sim} 10^{-2}.
\label{x3}
\end{equation}
We assume that
\begin{equation}
m_{\nu_e}, m_{\nu_{\mu}}, m_{\nu_s} < m_{\nu_{\tau}},
\label{x4}
\end{equation} 
and that the $\nu_{\tau}$ oscillates with the $\nu_s$ with small 
mixing, i.e. $ \sin^2 2\theta_0^{\tau s} \ll 1$.
In fact it turns out that the above 
assumptions are actually necessary if $N^{BBN}_{eff} = 4$ 
is to be avoided.

Of course there will be many
specific particle physics models consistent with the
assumptions, Eqs.(\ref{x3},\ref{x4}).
For example, Ref.\cite{ls} discusses one such model where
in addition to approximately maximal $\nu_\mu \to \nu_s$
oscillations there are $\nu_e \to \nu_\mu,\nu_s$
oscillations with parameters consistent
with the small angle MSW solution of the solar neutrino 
problem ($\sin^2 2\theta_0 \sim 10^{-2}$, 
$|\delta m^2| \sim 10^{-5}\ eV^2 $). 
Alternatively it is also possible that $\nu_e \to \nu_\mu$ 
oscillate in the region of parameter space suggested by the LSND
experiment\cite{lsnd}. Note however that models with
3 ordinary and 3 sterile neutrinos may be
quantitatively different due to the additional oscillation
modes possible. 
We intend to discuss some of these models in a 
forthcoming paper\cite{fv5}.

In this four neutrino scenario $\nu_\tau$ is assumed to be the 
heaviest neutrino so that $\nu_\tau \to \nu_s$ oscillations have 
$\delta m^2_{\tau s} < 0$, and create $L_{\nu_\tau}$ first at 
a temperature 
\begin{equation}
T_c^{\tau s} \approx 
16\left({-\delta m^2_{\tau s}\over eV^2}\right)^{1\over 6}
\ MeV.
\label{dons2}
\end{equation}
Note that the creation of $L_{\nu_\tau}$ implies that the matter
term for $\nu_\mu \to \nu_s$ oscillations is also generated
(note that the $a_{\mu s}(p)$ term in Eq.(\ref{sal}) is proportional to
$L^{(\mu)} \simeq 2L_{\nu_\mu} + L_{\nu_\tau} + L_{\nu_e}$).
Thus, the creation of a large $L_{\nu_\tau}$ asymmetry
also implies the creation of a large $L^{(\mu)}$ function
which can potentially suppress the $\nu_\mu \to \nu_s$ oscillations.
For example, for maximal vacuum mixing, the matter mixing angle 
for $\nu_\mu \to \nu_s$ oscillations is, 
\begin{equation}
\sin^2 2\theta_m^{\mu s} = 
{1 \over 1 + (a_{\mu s} \mp b_{\mu s})^2}.
\end{equation}
In order to simply explain the qualitative behaviour of this
oscillation system it is useful to consider the
quantities $\langle a_{\alpha s} \rangle$,
where $\langle a_{\alpha s} \rangle \equiv 
a_{\alpha s} (p=3.15T)$
and $p \simeq 3.15T$ is the mean momentum of a neutrino in a Fermi
Dirac distribution with zero chemical potential. 
[Of course in our numerical work the momentum distribution will
be taken into account].  In the region $T \sim T_c^{\tau s}$ where 
the exponential growth of $L_{\nu_\tau}$ occurs, $L_{\nu_\tau}$ 
is generated so that $\langle a_{\tau s} \rangle \sim 10$.
If $\nu_\mu \to \nu_s$ oscillations do not create significant 
$L_{\nu_\mu}$, then 
\begin{equation}
|\langle a_{\mu s} \rangle|
\simeq {1 \over 2}{|\delta m^2_{\tau s}| \over 
|\delta m^2_{\mu s}|}
|\langle a_{\tau s} \rangle| \gg 1 
\ if \ |\delta m^2_{\mu s}| \ll |\delta m^2_{\tau s}|.
\end{equation}
Thus, if $\nu_\mu \to \nu_s$ oscillations do not have
time to generate significant $L_{\nu_\mu}$ then the large
matter term $a_{\mu s}$ will be generated which will
suppress the $\nu_\mu \to \nu_s$ oscillations.
During the subsequent evolution, $\langle a_{\tau s} \rangle 
\sim 1$, so that $\langle a_{\mu s} \rangle $ remains very 
large and thus the $\nu_\mu \to \nu_s$ oscillations will
always be heavily suppressed (until very low temperatures
where $T \ll T^f_{\nu_\tau}$).  It
is important to realise though, that in response to the
$L_{\nu_\tau}$ created by $\nu_\tau \to \nu_s$ oscillations
the $\nu_\mu \to \nu_s$ oscillations can potentially
create $L_{\nu_\mu}$ such that $L^{(\mu)} \to 0$, since this
is an approximate fixed point of the $\nu_\mu \to \nu_s$ oscillation
system. Note that the evolution of $L_{\nu_\mu}$ due to 
$\nu_\mu \to \nu_s$ oscillations is dominated by the oscillations
in the resonance region where $b_{\mu s} \approx a_{\mu s}$
(assuming that $L_{\nu_\tau} > 0$ for definiteness)\footnote{
Of course if $L_{\nu_\tau} > 0$, then the $\nu_\mu \to \nu_s$
MSW resonance occurs for the neutrino oscillations
(rather than the anti-neutrino oscillations),
which create $L_{\nu_\mu} < 0$.}.  The resonance
momentum for maximal $\nu_\mu \to \nu_s$ oscillations
can be obtained from the condition $\lambda(p) =0$.
From Eqs.(\ref{sf},\ref{sal}) it follows that $p_{res}^{\mu s}$ 
is given by 
\begin{equation}
p_{res}^{\mu s} = {M_W^2 L^{(\mu)}\over T A_\mu}.
\end{equation}
Thus, $p_{res}^{\mu s}$
is proportional to $L^{(\mu)} \simeq 2L_{\nu_\mu}
+ L_{\nu_\tau} + L_{\nu_e}$.
As $L_{\nu_\tau}$ is created this causes $p_{res}^{\mu s}$ to
increase.  The $\nu_\mu \to \nu_s$ oscillations either create
$L_{\nu_\mu}$ fast enough so that $p_{res}^{\mu s} \stackrel{<}{\sim}
\langle p \rangle \simeq 3.15T$ or they do not.
If the $\nu_\mu \to \nu_s$ oscillations create $L_{\nu_\mu}$ 
sufficiently, then $\langle |a_{\mu s}-b_{\mu s}| \rangle 
\stackrel{<}{\sim} 10$.  Furthermore the $\nu_\mu \to \nu_s$ 
oscillations will continue to keep $|\langle a_{\mu s} 
- b_{\mu s}| \rangle \stackrel{<}{\sim} 10$ for lower temperatures 
because these oscillations become more effective at lower 
temperatures since they are not suppressed so much by the collisions.  
Thus, we need to compute the evolution of $L_{\nu_\tau},L_{\nu_\mu}$ 
due to $\nu_\tau \to \nu_s$ and $\nu_\mu \to \nu_s$ oscillations
through the high temperature region $({T_c^{\tau s}\over 2} 
\stackrel{<}{\sim} T \stackrel{<}{\sim} T_{initial}$)
in order to obtain the parameter space where 
$\nu_\tau \to \nu_s$ oscillations
generate a large $L^{(\mu)}$ which is {\it not} subsequently 
destroyed by $\nu_\mu \to \nu_s$ oscillations.
Of course the lower temperature 
($T \stackrel{<}{\sim} T_c^{\tau s}/2$) evolution of lepton 
number can also affect BBN in several ways, and we will 
return to this issue in section V.

As in Ref.\cite{fv2},
we assume that the $\nu_\tau, \nu_\mu, \nu_s$ system reduces to two 
two-flavour oscillations, $\nu_\tau \to \nu_s$ and $\nu_\mu \to
\nu_s$. This simplifying assumption is discussed in some
detail in Ref.\cite{bvw}.   
Heuristically, it is expected that
this simplifying assumption is justified
because the MSW resonance momentum of 
the two oscillation modes are generally different
(and different to the $\nu_\tau \to \nu_\mu$ resonance
momentum due to the matter effects). Also it
should be noted that the
$\nu_\tau \to \nu_\mu$, $\nu_\tau \to \nu_e$ and
$\nu_\mu \to \nu_e$ oscillations can be approximately
neglected at high
temperatures [Here `high' means $T \sim T_c^{\tau s}$
which is typically in the range $20 \stackrel{<}{\sim} 
T_c^{\tau s}/MeV
\stackrel{<}{\sim} 60$], because these oscillations do not 
create or destroy lepton number efficiently at high
temperature because there are almost equal numbers of
$\nu_e, \nu_\mu, \nu_\tau$ neutrinos.
Also $\nu_e \to \nu_s$ oscillations would be expected to be
heavily suppressed by the large tau lepton number (assuming
that $|\delta m^2_{e s}| \ll |\delta m^2_{\tau s}|$).

For the $\nu_\tau \to \nu_s$ and $\nu_\mu \to \nu_s$
oscillations we define a density matrix
(say ${\bf P}(p)$ for $\nu_\tau \to \nu_s$ oscillations 
and ${\bf Q}(p)$ for $\nu_\mu \to \nu_s$ oscillations). 
Thus, we have
\begin{eqnarray}
N_{\nu_\tau}(p) &=& {1 \over 2}P_0(p)\left(1 + P_z(p)\right)N^0(p),
\ N_{\nu_\mu}(p) = {1 \over 2}Q_0(p)\left(1 + Q_z(p)\right)N^0(p),
\nonumber \\
N_{\nu_s}(p) &=& {1 \over 2}P_0(p)\left(1 - P_z(p)\right)N^0(p) =
{1 \over 2}Q_0(p)\left(1 - Q_z(p)\right)N^0(p).
\end{eqnarray}
The evolution of the functions
$P_i(p), Q_i(p)$ are given by 
Eqs.(\ref{b1},\ref{b2}), where $\delta m^2, \sin^2 2\theta_0 
= \delta m^2_{\tau s}, \sin^2 2\theta_0^{\tau s}$ 
[$\delta m^2_{\mu s}, \sin^2 2\theta_0^{\mu s}$] in 
Eq.(\ref{sf}) for $P_i(p)$ [$Q_i(p)$]
(of course, for the evolution equations for
$P_i(p)$ [$Q_i(p)$], $\alpha = \tau$ [$\mu$] in Eq.(\ref{sal})).
At each time step, the lepton numbers $L_{\nu_{\tau}}, 
L_{\nu_{\mu}}$ are computed using Eq.(\ref{wed})
[in the case of $dL_{\nu_\mu}/dt$, $P_i(p) \to Q_i(p)$]. 
The two sets of evolution equations for $P_i (p), Q_i(p)$
are coupled together because they
each depend on both $L_{\nu_\tau} $ and $L_{\nu_{\mu}}$. 
They are also coupled together through the
population of the sterile state.
In particular the population of $\nu_s$
by $\nu_\tau \to \nu_s$ oscillations
can affect $\nu_\mu \to \nu_s$ oscillations.
Similarly the population of $\nu_s$ 
by $\nu_\mu \to \nu_s$ oscillations
can affect $\nu_\tau \to \nu_s$ oscillations.
This effect can only be important if the population
of sterile neutrinos becomes relatively large, 
i.e. $n_{\nu_s} \stackrel{>}{\sim} 0.1$.
Note that the population of $\nu_s$ due to 
$\nu_\mu \to \nu_s$ oscillations is negligible in the 
temperature region $T \sim T_c^{\tau s}$.  
This is because $T_c^{\tau s}$ is typically high
enough so that the $\nu_\mu \to \nu_s$ oscillations
are suppressed by the matter effects and also
damped by collisions so that they cannot populate
a significant number of sterile states.  However, if 
$\sin^2 2\theta_0^{\tau s}$ is large enough, then 
the $\nu_\tau \to \nu_s$ oscillations generate 
significant and
approximately equal numbers of  $\nu_s, \ \bar\nu_s$
in the region $T \stackrel{>}{\sim} T_c^{\tau s}$. 
This population of sterile neutrinos will obviously affect
the $\nu_\mu \to \nu_s$ oscillations.
The population of $\nu_s, \ \bar \nu_s$ by $\nu_\tau 
\to \nu_s$ oscillations can be incorporated by noting that 
\begin{equation}
{N_{\nu_s}(p) \over N^0(p)} =
{1 \over 2}P_0(p)(1 - P_z(p)) = {1 \over 2}Q_0 (p)(1 - Q_z(p)),
\end{equation}
and similarly for the anti-neutrinos.
This obviously implies that
the rate of change of $Q_i(p)$ due to $\nu_\tau \to \nu_s$
oscillations (we denote this quantity by
${\partial \over \partial t}Q_i(p)|_{\nu_\tau \to \nu_s}$) 
satisfies
\begin{equation}
{\partial \over \partial t}
\left[Q_0 (p)(1 - Q_z(p))\right]|_{\nu_\tau \to \nu_s} \simeq
{\partial \over \partial t} 
\left[P_0 (p)( 1 - P_z(p)) \right].
\label{uuu1}
\end{equation}
Furthermore, the population of sterile neutrinos by
$\nu_\tau \to \nu_s$ oscillations obviously does not directly
affect the number of $\nu_\mu$ neutrinos. Thus,
\begin{equation}
{\partial \over \partial t} 
N_{\nu_\mu}(p)|_{\nu_\tau \to \nu_s} = 
{\partial \over \partial t} 
\left[ Q_0 (p)(1 + Q_z(p)) \right]|_{\nu_\tau \to \nu_s} = 0.
\label{uuu2}
\end{equation}
Solving Eqs.(\ref{uuu1},\ref{uuu2}) (and the analogous 
equations for the anti-neutrinos) implies that
\begin{eqnarray}
{\partial \over \partial t}\bar Q_0(p)|_{\nu_\tau \to \nu_s} 
& = &
{\partial \over \partial t}Q_0(p)|_{\nu_\tau \to \nu_s}  
= {1 \over 2}
\left({\partial P_0(p) \over \partial t}(1 - P_z(p)) - {
\partial P_z(p) \over \partial t}P_0(p)\right),
\nonumber \\
{\partial \over \partial t}\bar Q_z(p)|_{\nu_\tau \to \nu_s} 
& = &
{\partial \over \partial t} Q_z(p)|_{\nu_\tau \to \nu_s} 
\nonumber \\
&=& - {1 \over 2} {(1 + Q_z(p)) \over Q_0(p)}
\left({\partial P_0(p) \over \partial t}(1 - P_z(p)) - {
\partial P_z(p) \over \partial t}P_0(p)\right),
\label{haa}
\end{eqnarray}
where here we have neglected the tiny difference in $\nu_s, \
\bar \nu_s$ populations (which is actually necessary for self 
consistency).  When solving the evolution equations for 
$P_i(p), Q_i(p)$ we have included the contribution, Eq.(\ref{haa}).   

Our initial conditions for the numerical integration
are (as explained in section III),
$P_{x,y}(p) = \bar P_{x,y}(p) = 0$
and $P_z(p) = \bar P_z(p) = P_0(p) = \bar P_0(p) = 1$
(and similarly for the $Q_i(p), \bar Q_i(p)$, $i=x,y,z,0$).
We also set the initial values of all of the
neutrino asymmetries to zero, and took $T_{initial} =
4T_{res}^{\tau s}$. Of course the results do not
depend on $T_{initial}$ so long as it is high enough
(and $T_{initial} = 4T_{res}^{\tau s}$ is certainly
high enough).  As we discussed in section III (and in Ref.
\cite{fv2}), the results 
are also independent of the initial values of the neutrino 
asymmetries so long as they are less than about $10^{-5}$.
We also utilise the approximation of integrating around
the region of the MSW resonance [taking $f=7$ in
Eq.(\ref{stu})] for both $\nu_{\tau} \to \nu_s$ and 
$\nu_{\mu} \to \nu_s$ oscillations.  We checked the stability 
of our results by taking larger slices of momentum space. 

Performing the necessary numerical work,  
we obtained the region of parameter space where
the $L^{(\mu)}$ created by $\nu_{\tau} \to \nu_s$
oscillations is not destroyed by maximal $\nu_\mu \to
\nu_s$ oscillations. This region
is given in Figure 2.
Of course, we must also require that the
sterile neutrinos do not become significantly
populated by $\nu_\tau \to \nu_s$ oscillations.
Recall from Eq.(\ref{large}) that
$\delta N_{eff}^{BBN} \stackrel{<}{\sim} 0.6$
implies the constraint\cite{fv2},
\begin{equation}
\sin^2 2\theta_0 \stackrel{<}{\sim} 4 \times 
10^{-5}\left[ {eV^2 \over |\delta m^2_{\tau s}|}
\right]^{1 \over 2}.
\label{yyy}
\end{equation}
Note that $\nu_\tau \to \nu_s$ oscillations populate 
the sterile neutrinos in the temperature region $
T \stackrel{>}{\sim} T_c^{\tau s}$ where $a_{\tau s}(p)$
is very small. The constraint, Eq.(\ref{yyy}) is given
by the dashed-dotted line in Figure 2. There are also 
other contributions to $N_{eff}^{BBN}$, which are due to 
the low temperature evolution of $L_{\nu_\tau}, L_{\nu_\mu}$ 
and $L_{\nu_e}$, including the population of sterile neutrinos
at low temperatures.  We will return to these issues in section V.

Let us now compare our new results with the earlier calculations 
of Ref.\cite{fv2}.  In Figure 7 of Ref.\cite{fv2} we took 
$|\delta m^2_{atmos}| = 10^{-2}\ eV^2$
and computed the allowed region by numerically
integrating an approximate
solution of the quantum kinetic equations.
This approximation (which we called the `static approximation'
in Ref.\cite{fv2}) holds provided that the system is 
smooth on the interaction time scale $1/D(p)$.
For this $\delta m^2_{atmos}$ our new results are in
agreement with the previous calculation of Ref.\cite{fv2}
in the region $\sin^2 2\theta_0 \stackrel{<}{\sim} 
3 \times 10^{-7}$. This result was anticipated in Ref.\cite{fv2},
where it was found that the evolution of lepton number
was smooth enough for the static approximation to be acceptable
in this parameter range.  In the case where 
$\sin^2 2\theta_0 \stackrel{>}{\sim} 3 \times 10^{-7}$
our new results show that the allowed region is somewhat 
smaller when compared with the approximation of Ref.\cite{fv2}.

Finally recall that in the context of the
standard big bang model, the age of the 
Universe implies an upper bound on the mass
of the tau neutrino which is\cite{subir} 
\begin{equation}
m_{\nu_\tau} \stackrel{<}{\sim} 100 \ eV,
\end{equation}
and hence
\begin{equation}
|\delta m^2_{\tau s}| \stackrel{<}{\sim} 10^4 \ eV^2
\label{ddfe}
\end{equation}
Some authors (including myself) have
taken a more stringent bound of $m_{\nu_\tau} \stackrel{<}{\sim}
40 \ eV$ (which implies $|\delta m^2_{\tau s}|
\stackrel{<}{\sim} 10^3 \ eV^2$).
While the latter bound certainly appears to be favoured,
we feel it is prudent to be cautious,
and we therefore adopt the weaker bound Eq.(\ref{ddfe}).

\vskip 0.5cm
\noindent
{\bf V Detailed implications for big bang nucleosynthesis}
\vskip 0.5cm
Let us briefly summarise the story so far. In the 
region above the solid 
line(s) in Figure 2, $\nu_\tau \to \nu_s$ 
oscillations create a large
$L_{\nu_\tau}$ which suppresses $\nu_\mu \to \nu_s$
oscillations so that these oscillations cannot
significantly populate the sterile state (for
$T \stackrel{>}{\sim} 0.5 \ MeV$).
Let us denote the contribution to $N^{BBN}_{eff}$ from
the oscillations,
$\nu_{\mu} \to \nu_s$ ($\nu_\tau \to \nu_s$ in the region
$T \stackrel{>}{\sim} T_c^{\tau s}$) 
by the notation $\delta_1 N^{BBN}_{eff}$
($\delta_2 N^{BBN}_{eff}$).  With this notation, 
in the parameter region to the right (left)
of the dashed-dotted line $\delta_2 N^{BBN}_{eff} > 0.6$
($\delta_2 N^{BBN}_{eff} < 0.6$).
Also, in the region above (below) the solid line(s)
in Figure 2, $\delta_1 N^{BBN}_{eff}
\ll 0.1$ ($\delta_1 N^{BBN}_{eff} + \delta_2
N^{BBN}_{eff} \simeq 1$).

It is important to realise that
$N_{eff}^{BBN}$ depends on the number densities
of $\nu_e, \bar \nu_e$  through
BBN nuclear reaction rates (as well as the expansion rate).
Thus, it is important to study the evolution of the four
neutrino system down to low temperatures $T \sim 0.5$ MeV.
The evolution of this four neutrino system down to low
temperatures has already been studied in some detail
in Ref.\cite{fv3} and we include 
a discussion here for completeness.  Since the final value of 
$L_{\nu_\tau}$, $L^f_{\nu_\tau}$, is quite big there is
a significant modification to the momentum distribution
of the tau and anti-tau neutrinos. If the tau neutrinos
also oscillate into muon and electron neutrinos, then some
of the tau lepton number can be transferred to the electron and
muon neutrinos.  In other words small $L_{\nu_e}, L_{\nu_{\mu}}$
will be created. The details are quite independent of 
the intergenerational mixing angles so long as they are small 
(here, small means $\sin^2 2\theta_0 \stackrel{<}{\sim} 
0.1$)\footnote{They cannot be arbitrarily small lest the 
oscillations become non adiabatic (typically they must
be greater than about $3 \times 10^{-10}$).} and depend only
on $\delta m^2$. In Ref.\cite{fv3} the contribution to 
$N_{eff}^{BBN}$ due to the modification of the $\nu_e, 
\bar \nu_e$ momentum distributions (which we denote by 
$\delta_3 N^{BBN}_{eff}$) was found to be
\begin{eqnarray}
\delta_3 N_{eff}^{BBN} &\simeq & 0 (0), \ for \  
|\delta m^2|/eV^2 \stackrel{<}{\sim} 3, \nonumber \\
\delta_3 N_{eff}^{BBN} &\simeq & -0.45 (0.45) \ for \ 3 
\stackrel{<}{\sim} |\delta m^2|/eV^2 \stackrel{<}{\sim} 1000, \nonumber
\\ \delta_3 N_{eff}^{BBN} &\simeq & -0.50 (0.50)  \ for \
|\delta m^2|/eV^2 \stackrel{>}{\sim} 1000, 
\label{kj}
\end{eqnarray}
for $L_{\nu_e} > 0$ ($L_{\nu_e} < 0$).
Note that $\delta_3 N_{eff}^{BBN}$
in the above equation is actually continuous so there is
really a smooth transition region between the three $\delta m^2$
regions which we have neglected for simplicity\footnote{
Actually it should be noted that $\delta_2 N_{eff}^{BBN},
\delta_3 N_{eff}^{BBN}$ and $\delta_4 N_{eff}^{BBN}$
are all continuous above the solid line(s) in Figure 2.
This means that their values change smoothly. The
contribution $\delta_1 N_{eff}^{BBN}$, on the other hand
is discontinuous across the solid line(s).}.
The creation of the lepton number can also change 
the expansion rate since the energy density of the Universe 
will be modified a bit by the modification of the number 
and momentum distributions that occur when the large 
$L_{\nu_\tau}$ neutrino asymmetry
is created. The contribution to $N_{eff}^{BBN}$ due 
to this effect (which we denote by $\delta_4 N^{BBN}_{eff}$)
was calculated to be\cite{fv3},
\begin{eqnarray}
\delta_4 N_{eff}^{BBN} &\simeq & 0 \ for \  
|\delta m^2|/eV^2 \stackrel{<}{\sim} 3, \nonumber
\\ \delta_4 N_{eff}^{BBN} &\simeq & -0.05 \ for \ 3 
\stackrel{<}{\sim} |\delta m^2|/eV^2 \stackrel{<}{\sim} 
1000, \nonumber \\ \delta_4 N_{eff}^{BBN} &\simeq & 0.40   \ for \
|\delta m^2|/eV^2 \stackrel{>}{\sim} 1000. 
\label{kjj}
\end{eqnarray}
Finally note that the results in Eq.(\ref{kj}) and Eq.(\ref{kjj})
are only valid provided that $\delta_1 N_{eff}^{BBN}
+ \delta_2 N_{eff}^{BBN} \stackrel{<}{\sim} 0.1$.
This is only approximately true in the region above
the solid line(s) in Figure 2 and in the region where 
$\nu_\tau \to \nu_s$ oscillations do not significantly populate
$\nu_s$ at high temperature $T \stackrel{>}{\sim} T_c$.
The region of parameter space where $\nu_\tau \to \nu_s$ 
oscillations do not significantly populate $\nu_s$ at high temperature
(i.e. $\delta_2 N_{eff}^{BBN} \stackrel{<}{\sim} 0.1$) is given by
\footnote{
Note that there is a mistake in the
corresponding equation in Ref.\cite{fv3}. Eq.(\ref{lar})
is the correct equation.},
\begin{equation}
\sin^2 2\theta_0 \stackrel{<}{\sim} 
1.3 \times 10^{-5}\left[{eV^2 \over |\delta m^2|}\right]^{1
\over 2}.
\label{lar}
\end{equation}
Comparing this equation with Eq.(\ref{large}) (which assumes 
that $\delta_2 N_{eff}^{BBN} \stackrel{<}{\sim} 0.6$) we 
see that it is only slightly more stringent.
If there is a significant number density of $\nu_s$ coming
from $\nu_\tau \to \nu_s$ oscillations at high temperature,
then this will reduce the size of $L^f_{\nu_\tau}$. The effects
given in Eq.(\ref{kj}) and Eq.(\ref{kjj}) will therefore
be reduced. 

Observe that in the region in Figure 2
below the solid line(s), the sterile neutrino will be
populated by $\nu_\mu \to \nu_s$ oscillations
at a temperature $T \sim 6-10$ MeV.
When this occurs the growth in $L_{\nu_\tau}$ is cutoff 
because the number of tau and sterile neutrinos become
approximately equal.  A these temperatures, $|L_{\nu_\tau}| 
\stackrel{<}{\sim} 10^{-2}$ so that the possible modifications 
of $\nu_e, \bar \nu_e$ are negligible in this case.
In other words $N_{eff}^{BBN} \simeq 4,$
with $\delta_1 N_{eff}^{BBN} + \delta_2 N_{eff}^{BBN}
\simeq 1, \ \delta_3 N_{eff}^{BBN}, \ \delta_4 N_{eff}^{BBN}
\ll 0.1.$ 

To summarise, it is possible to identify 4 distinct 
contributions to $N_{eff}^{BBN}$. They are: 
\vskip 0.3cm
\noindent
(1) $\delta_1 N_{eff}^{BBN}$. This is the contribution 
to $N_{eff}^{BBN}$ which arises from the change
in the expansion rate of the Universe due
to the population of the sterile neutrinos by
$\nu_\mu \to \nu_s$ oscillations.
\vskip 0.4cm
\noindent
(2) $\delta_2 N_{eff}^{BBN}$. This is the contribution
to $N_{eff}^{BBN}$
which arises from the change in the expansion rate
due to the population of sterile neutrinos by
$\nu_\tau \to \nu_s$ oscillations 
for temperatures before the exponential growth of 
the $L_{\nu_\tau}$ occurs (i.e. $T \stackrel{>}{\sim}
T_c^{\tau s}$).
\vskip 0.4cm
\noindent
(3) $\delta_3 N_{eff}^{BBN}$. This is the contribution
to $N_{eff}^{BBN}$ from the direct modification 
of the nuclear reaction rates (such as 
$\nu_e + N \leftrightarrow P + e^-$, 
$\bar \nu_e + P \leftrightarrow N + e^+$).
This occurs because of the modification  
of the momentum distributions of $\nu_e, \bar \nu_e$ 
due to a small $L_{\nu_e}$ asymmetry which
is transferred from $L_{\nu_\tau}$ by $\nu_\tau
\to \nu_e$ oscillations.
\vskip 0.4cm
\noindent
(4) $\delta_4 N_{eff}^{BBN}$. This is the contribution
to $N_{eff}^{BBN}$ from the change in the expansion rate
due to the modification of the energy densities
of the neutrinos at low temperature
$T \sim T^f_{\tau s}$ which is caused by the large
$L_{\nu_\tau}$.
\vskip 0.3cm
\noindent
Note that both the effects (2) and (4) are essentially
due to the population of $\nu_s$ by $\nu_\tau \to \nu_s$
oscillations. We label them distinctly because
the effect (2) occurs at high temperatures
before the generation of neutrino asymmetry occurs
(in this region $\nu_s, \bar \nu_s$ are populated
approximately in equal numbers) while
the effect (4) occurs because of the population
of $\bar \nu_s$ (taking $L_{\nu_\tau} > 0$ for
definiteness) which necessarily occurs in the
temperature region where the neutrino asymmetry
nears its final value.  Finally, the effective neutrino 
number for BBN is related to the 
contributions, $\delta_i N_{eff}^{BBN}$, 
in the obvious way
\begin{equation}
N_{eff}^{BBN} = 3 + \sum^4_{i=1} \delta_i N_{eff}^{BBN}.
\end{equation}

The results of this section are summarised in Figures 3,4.
Figure 3 (Figure 4) is for the case where $L_{\nu_e} > 0$ 
($L_{\nu_e} < 0$)\footnote{
The sign of $L_{\nu_e}$ is the same as the
sign of $L_{\nu_{\tau}}$, which cannot be
predicted at the moment (see Ref.\cite{fv2} for
some discussion on this point).}.
As these figures show, there is a significant chunk of 
parameter space where $N^{BBN}_{eff} < 3$.  Thus, not only
is the $\nu_\mu \to \nu_s$ oscillation solution to the 
atmospheric neutrino anomaly consistent with a stringent
BBN bound of $N_{eff}^{BBN} < 3.6$, but there is a 
range of parameters where this solution implies that
{\it there are less than 3 neutrinos for BBN}.
It is interesting that there are some recent indications that
$N^{BBN}_{eff} < 3$ is actually favoured by the data
especially if the low deuterium abundance measurements
$D/H \sim 3 \times 10^{-5}$\cite{eee}
in quasar absorption systems are correct\cite{lang,oth}
\footnote{It has been argued that $N_{eff}^{BBN}$ can be as 
large as $4.5$\cite{sar} and still be consistent with the
data. However this paper assumed that $D/H$ could
be as large as $2.5 \times 10^{-4}$ which seems
to be disfavoured by recent studies\cite{eee}
(see also Ref.\cite{ts} for a review). Nevertheless,
things are not completely clear at the moment, so
one should keep in mind the possibility that
$N_{eff}^{BBN}$ as large as $4.5$, as suggested
by Ref.\cite{sar}, may be consistent with the standard
big bang model.}.
The lesson here is that there are well motivated extensions
of the standard model which give $N_{eff}^{BBN} < 3$.
Thus, if it turns out that $N_{eff}^{BBN} < 3$ 
is required for consistency of the standard
big bang model, then this
should be taken seriously as an indication that sterile
neutrinos exist\footnote{
Of course neutrino oscillations is not the only 
mechanism which can give $N^{BBN}_{eff} < 3$, however
it is perhaps the most well motivated given the 
existing neutrino anomalies.
For some other ways to get $N^{BBN}_{eff} < 3$,
see for example, Ref.\cite{dd}.}.

Finally, the scenario we have given in this section holds
provided that $\nu_e \to \nu_s$ oscillations can be
neglected. In the case where $|\delta m^2_{es}|
\ll 1 \ eV^2$, as happens for the particular
case studied in Ref.\cite{ls} (where
$\nu_e \to \nu_\mu, \nu_s$ oscillations
solve the solar neutrino problem), there is no impact for
BBN from $\nu_e \to \nu_s$ oscillations.
However, in the case where $|\delta m^2_{es}|/eV^2
\stackrel{>}{\sim} 1$
\footnote{Note that from Eq.(\ref{x3})
it follows that $\nu_\mu$ and $\nu_s$
are maximal mixtures of two nearly
degenerate mass eigenstates in this case.}
(which is compatible with the LSND
experiment), it turns out that the $\nu_e \to \nu_s$
oscillations cannot be neglected for BBN.
In particular, consider the possibility that 
\begin{equation}
m_{\nu_e} < m_{\nu_\mu} \approx m_{\nu_s} < m_{\nu_\tau} \
\ and \ \delta m^2_{es} \sim 1 \ eV^2,
\end{equation}
along with the requirement that the atmospheric neutrino
anomaly is solved by $\nu_\mu \to \nu_s$ oscillations,
Eq.(\ref{x3}).
We will assume that $\sin^2 2\theta_0^{es}$ is small
enough so that $\nu_e \to \nu_s$ oscillations
are unable to create significant $L_{\nu_e}$ at
high temperatures, $T \sim T_c^{\tau s}$.
[The situation at high temperature 
is qualitatively similar to the case studied
in section IV].
The low temperature evolution of $L_{\nu_e}$
will be dominated by the MSW transitions.
It is instructive to consider the resonance
momentum of $\nu_e \to \nu_s$ oscillations,
$p_{res}^{es}$. At low temperatures, where
the $b(p)$ terms can be neglected, the resonance condition
arises from the equation $\pm a(p) = \cos2\theta_0$ ($\simeq 1$
in this case for both $\nu_\tau \to \nu_s$ and $\nu_e \to 
\nu_s$ oscillations).
Taking $L_{\nu_\tau} > 0$ for definiteness, which means that
at low temperatures, the $\bar \nu_\tau \to \bar \nu_s$
have a MSW resonance, and the $\nu_e \to \nu_s$
oscillations have a MSW resonance
(rather than the $\bar \nu_e \to \bar \nu_s$ oscillations).
This opposite behaviour occurs because 
$\delta m^2_{es} > 0$ while $\delta m^2_{\tau s} < 0$.
Thus it is clear that the $\nu_e \to \nu_s$
oscillations generate $L_{\nu_e}$ with the
opposite sign to $L_{\nu_\tau}$.
Using Eq.(\ref{sal}),
\begin{equation}
{p_{res}^{es} \over T} \simeq {\pi^2 \delta m^2_{es} \over 
4\zeta (3) \sqrt{2} G_F T^4 L^{(e)}}
\approx 0.5{\delta m^2_{es} \over eV^2}{MeV^4 \over T^4}
{0.23 \over L^{(e)}}.
\label{ww}
\end{equation}
Eventually, for $T \stackrel{<}{\sim} T^f_{\nu_\tau}$ [see
Eq.(\ref{nau})],
$L_{\nu_\tau} \to L^f_{\nu_\tau} \simeq
0.23$ (for $3 \stackrel{<}{\sim} |\delta m^2_{\tau s}|/eV^2
\stackrel{<}{\sim} 1000$).
For these temperatures, $L^{(e)} \approx 2L_{\nu_e} + 
L^f_{\nu_\tau}$.
Thus, as $T \to 0$, Eq.(\ref{ww}) 
implies that $p_{res}^{es}/T \to \infty$ and
there will be significant MSW conversion of $\nu_e 
\to \nu_s$. In this case there cannot be complete
MSW conversion because of the 
structure of Eq.(\ref{ww}). In fact Eq.(\ref{ww}),
implies that the creation of $L_{\nu_e}$ actually increases
the rate at which $p_{res}^{es}/T$ moves
through the momentum distribution. Thus, 
$L_{\nu_e}$ is created much more rapidly than $L_{\nu_\tau}$.
Indeed it must be created so rapidly that the oscillations
are not completely adiabatic which is why
incomplete MSW conversion occurs in this case.
In fact, it is possible to show
that as $p_{res}^{es}/T \to \infty$,
$L^{(e)} \to 0$ (approximately), i.e. $L_{\nu_e}
\to -L^f_{\nu_\tau}/2 \simeq -0.12$.
Since significant generation of $L_{\nu_e}$ cannot
occur until temperatures where $p_{res}^{es}/T
\stackrel{>}{\sim} 1$, it follows from Eq.(\ref{ww})
that for $\delta m^2_{es} \ll 1 eV^2$ the
significant creation of $L_{\nu_e}$ occurs at
temperatures much less than $0.5$ MeV, so that there is
no significant impact for BBN in this case.
However, for $\delta m^2_{es} \stackrel{>}{\sim} 1\
eV^2$, the $L_{\nu_e}$ would be created
during the time when the $N + \nu_e \leftrightarrow P + e^-$,
$P + \bar \nu_e \leftrightarrow N + e^+$
reactions are still occurring.
Thus, in this case there will be a significant modification
to BBN.  This connection between the oscillations in the 
LSND parameter range and BBN is extremely 
interesting, but we will not present any quantitative
numerical results here. However, we do intend to
discuss this effect more fully in the context of models 
with 3 ordinary and 3 sterile neutrinos in the
forthcoming paper\cite{fv5}.

\vskip 0.5cm
\noindent
{\bf VI Implications for the hot+cold dark matter model}
\vskip 0.5cm
In the scenario considered in this paper,
where $\nu_\mu \to \nu_s$ oscillations solve the
atmospheric neutrino anomaly,
a BBN bound of $N_{eff}^{BBN} < 3.6$ implies
$m_{\nu_\tau} \stackrel{>}{\sim} 4\ eV$ for
$|\delta m^2_{atmos}| \simeq 10^{-2.5} \ eV^2$
(see Figure 2).  Neutrino masses in the eV range have 
long been considered to be interesting for cosmology.
This is because these neutrinos can make a significant
contribution to the energy density of the Universe.
In the standard big bang model, the contribution of
massive neutrinos to the energy density is given by
the well known formula,
\begin{equation}
\Omega_{neutrino} = {\sum_\alpha m_{\nu_\alpha}
\over h^2 92\ eV},
\end{equation}
where $h$ is the usual cosmological parameter 
parameterising the uncertainty in the Hubble constant.
Thus, neutrinos in the eV mass range are a well known 
and well motivated candidate for hot dark matter. 
Some studies have suggested that a hot+cold dark matter 
mixture with $\Omega_{neutrinos} \simeq 0.20 - 0.25$  
can nicely explain the structure formation\cite{sch}.  
If only one eV neutrino is assumed, which we take to be the
tau neutrino, then the neutrino mass favoured in the 
usual hot+cold dark matter scenario is\cite{sch} 
\begin{equation}
3 \ eV \stackrel{<}{\sim} m_{\nu_\tau} \stackrel{<}{\sim}
7 \ eV.
\end{equation}
This situation is modified somewhat when light sterile neutrinos
exist. The reason is that the $\nu_\tau \to \nu_s$ oscillations
generate such a large $L_{\nu_{\tau}}$
that the total number of tau neutrinos is actually significantly 
reduced. In the region where 
$3 \stackrel{<}{\sim} |\delta m^2_{\tau s}|/eV^2
\stackrel{<}{\sim} 1000$,
the $L^f_\nu \simeq 0.23$ [from Eq.(\ref{go})].
The large final lepton number occurs because
about $70\%$ of the anti-neutrinos have
been depleted (assuming $L_{\nu_\tau} > 0$) 
which means that the total number of tau neutrinos (i.e.
tau + anti -tau neutrinos) is approximately $0.65$ of 
a standard neutrino species.
(Note that the total number of neutrinos hasn't changed much,
the missing heavy anti-tau neutrinos have just been
converted into light sterile states.)
This significant depletion of tau neutrinos implies that
the tau neutrino mass favoured in the hot+cold dark matter 
scenario (which suggests that $\Omega_{neutrinos} 
\simeq 0.20 - 0.25$)  is actually about $50\%$ 
larger than the naive expectation.  Thus when $\nu_\tau$ 
oscillates with a light $\nu_s$, the hot+cold dark 
matter scenario actually favours a tau neutrino mass of 
\begin{equation}
5 \ eV \stackrel{<}{\sim} m_{\nu_\tau} \stackrel{<}{\sim}
10 \ eV,
\end{equation}
rather than $3 - 7$ eV.  This means that 
$|\delta m^2_{\tau s}|$ is expected to be in the range
\begin{equation}
25\ eV^2 \stackrel{<}{\sim} |\delta m^2_{\tau s}| \stackrel{<}{\sim} 
100 \ eV^2,
\label{reg}
\end{equation}
assuming here that the sterile neutrino is much lighter than the tau
neutrino.  The hot+cold dark matter region, Eq.(\ref{reg}) 
is the shaded band on Figure 2.
From this figure, we see that for $|\delta m^2_{atmos}| = 10^{-2} \
eV^2$ there is only a relatively small
range of $\sin^2 2\theta_0^{\tau s}$ where
the favoured hot+cold dark matter region is compatible with
$N_{eff}^{BBN} \stackrel{<}{\sim} 3.6$.
However, for lower values of $|\delta m^2_{atmos}|$ which are actually
favoured by the superKamiokande data, there is a significant 
range for $\sin^2 2\theta_0^{\tau s}$ 
where the favoured hot+cold dark matter region is compatible with
$N_{eff}^{BBN} \stackrel{<}{\sim} 3.6$.

Finally note that detailed studies\cite{sch} of structure
formation in the hot+cold dark matter model show that they are
sensitive to the number of eV neutrinos (not just 
$\Omega_{neutrinos}$). These studies typically assume that
the number of eV neutrinos (usually taken to be
degenerate) at the epoch of matter radiation
equality (we will denote this quantity by $N_{eff}^{dm}$)
is an integer, i.e.,  $N_{eff}^{dm} = 1,2,3$.
It is important to understand that this is only true
provided that sterile neutrinos do not exist.
Indeed, as we have just explained above,
we expect $N_{eff}^{dm} \simeq 0.65$ in the case where
only $\nu_\tau$ is in the eV mass range and oscillates
with a light sterile neutrino\footnote{
In Ref.\cite{bfv}, the 4 neutrino
model of Ref.\cite{vm} where $\nu_\mu \to \nu_\tau$ 
oscillations solved the atmospheric neutrino anomaly
was studied.  In that particular model $\nu_\mu$, $\nu_\tau$ 
are assumed to be approximately degenerate and in the eV 
mass range. In that model the oscillations of the
$\nu_{\mu} \to \nu_s$ and $\nu_\tau \to \nu_s$ 
created a large $L_\mu \simeq L_\tau \simeq 0.16$ and
$N_{eff}^{dm} \simeq 1.5$.}.

\vskip 1.5cm
\noindent
{\bf VII Implications for the comic microwave background}
\vskip 0.5cm
During the next decade or so, high precision measurements of the 
anisotropy of the cosmic microwave background (CMB)
will be performed by several experiments (such as the 
PLANK and MAP missions).
From these experiments it may be possible to estimate 
accurately the radiation content of the Universe at the
epoch of photon - matter decoupling\cite{turn}.
In this context it is important to note that sterile
neutrinos can leave their `imprint' on the cosmic
microwave background\cite{hr}.
Thus it is useful to introduce the quantities 
$N_{eff}^{l\nu}, N_{eff}^{h\nu}$
where $N_{eff}^{l\nu}$ ($N_{eff}^{h\nu}$) is the effective 
number of light neutrinos (heavy neutrinos) at the
epoch of photon decoupling. In this context, `light'  means 
much less than about an eV (which will make them 
relativistic at the epoch of photon decoupling) 
and `heavy' means more than about an eV (which will make them 
approximately non-relativistic). 
Of course in the minimal standard model of particle physics
which has 3 massless neutrinos, $N^{BBN}_{eff}
= N^{l\nu}_{eff} = 3, N^{h\nu}_{eff} = 0$.
However in the four neutrino model case that we are discussing
in this paper,
$N^{BBN}_{eff} \neq N_{eff}^{l\nu}$ and $N_{eff}^{h\nu} \neq 0$
in general.
[Note that $N_{eff}^{BBN} \neq N_{eff}^{l\nu}$ not just because of the
heavy tau neutrino but also because of the contribution of $L_{\nu_e}$
to $N_{eff}^{BBN}$].

If we assume that $m_{\nu_{\tau}} \stackrel{>}{\sim} 4 \ eV$ (as
suggested by Figure 2) and
$m_{\nu_e}, m_{\nu_\mu}, m_{\nu_s} \ll 1 eV$ then
\begin{equation}
N_{eff}^{h\nu} \simeq {n_{\nu_\tau} \over n_0},\
N_{eff}^{l\nu} = {\rho_{\nu_e} + \rho_{\nu_\mu} + 
\rho_{\nu_s} \over \rho_0},
\end{equation} 
where $n_i$ ($\rho_i$) are the mass (energy) densities with 
$n_0$ ($\rho_0$) being the mass (energy) density of a massless 
neutrino with Fermi-Dirac distribution with zero chemical potential.
Assuming that $3 \stackrel{<}{\sim} 
|\delta m^2_{\tau s}|/eV^2 \stackrel{<}{\sim} 1000$, 
using results from Ref.\cite{fv3},
\begin{equation}
N_{eff}^{h\nu} \simeq N_{eff}^{dm} \simeq 0.65,\
N_{eff}^{l\nu} \simeq 2.2.
\end{equation}
The precise measurements of the CMB may well prove to 
be quite useful in distinguishing between various competing
explanations of the neutrino anomalies, since each
model should leave quite a distinctive imprint on the CMB. 

\vskip 0.5cm
\noindent
{\bf VIII Concluding remarks}
\vskip 0.5cm
In this paper we have numerically integrated the
quantum kinetic equations to obtain the region of
parameter space where the $\nu_\mu \to \nu_s$ oscillation
solution to the atmospheric neutrino anomaly is
consistent with a BBN bound of 3.6 effective neutrinos. 
The consistency occurs because the $\nu_\tau \to \nu_s$
oscillations create a large $L_{\nu_\tau}$ asymmetry. This
large asymmetry modifies the effective potential for $\nu_\mu \to \nu_s$
oscillations which become heavily suppressed for the range
of parameters given in Figure 2.
These dynamically induced matter effects
prevent the $\nu_\mu \to \nu_s$ 
oscillations from significantly populating the sterile state for
the entire `allowed region' of Figure 2.
This work confirms and improves the previous study\cite{fv2}
which utilised an approximate solution of the
quantum kinetic equations (which we called the `static' approximation
in Ref.\cite{fv2}). We have also discussed the detailed
implications of this scenario for BBN (see figures 3,4), 
the hot+cold dark matter model, and also for the forthcoming 
precision measurements of the cosmic microwave background.

In this paper we assumed the existence of only one 
light sterile neutrino.  Probably 
the simplest four neutrino scheme which
can solve the atmospheric neutrino anomaly by $\nu_\mu 
\to \nu_s$ oscillations and also solve the solar
neutrino problem is the case considered in Ref.\cite{ls} 
(where small angle MSW enhanced $\nu_e \to \nu_\mu, \nu_s$
oscillations solve the solar neutrino problem).
Our results indicate that the tau neutrino mass should 
be larger than $3-4$ eV if consistency with a stringent
BBN bound of $3.6$ neutrinos is required.
Note however that the scenario of Ref.\cite{ls} cannot explain 
the LSND anomaly\cite{lsnd}.
The LSND experiment\cite{lsnd} has provided strong evidence 
for $\bar \nu_\mu \to \bar \nu_e$ and $\nu_\mu \to
\nu_e$ oscillations. The suggested parameter range
is $\sin^2 2\theta_0 \sim 10^{-2}$ and
\begin{equation}
0.2 \stackrel{<}{\sim} |\delta m^2_{lsnd}|/eV^2 
\stackrel{<}{\sim} 6.
\label{who}
\end{equation}
One may well wonder whether
or not there exists any four neutrino scheme which
can solve all three experimental anomalies
(with the $\nu_\mu \to \nu_s$ oscillations solving
the atmospheric neutrino anomaly), and still be 
consistent with a stringent BBN bound of $3.6$ neutrinos?
The only potential scheme which comes
to mind is\cite{bp}\footnote{Please excuse the sloppy notation
in Eq.(\ref{how}).  Of course the mass eigenstates are linear 
combinations of weak eigenstates.}
\begin{equation}
m_{\nu_\tau} \simeq m_{\nu_e} > m_{\nu_\mu}, m_{\nu_s},
\label{how}
\end{equation} 
with $\nu_e \to \nu_\tau$ oscillations solving the solar
neutrino problem, either through vacuum oscillations 
or through MSW enhanced oscillations (this means that
$|\delta m^2_{\tau e}| \stackrel{<}{\sim} 10^{-3}\ eV^2$).
The LSND anomaly is explained by $\nu_\mu \to \nu_e$ oscillations
with
\begin{equation}
\delta m^2_{\tau s} \simeq \delta m^2_{es} \simeq 
\delta m^2_{e\mu} \simeq \delta m^2_{lsnd} \stackrel{<}{\sim}
6 \ eV^2.
\end{equation}
In this scenario, $\nu_\tau \to \nu_s$ oscillations will
generate $L_{\nu_\tau}$  and $\nu_e \to \nu_s$ oscillations
generate $L_{\nu_e}$.
However because the $|\delta m^2_{lsnd}|$ is so low,
the results in Figure 2 indicate that,
except for a very small region of parameter space
($|\delta m^2_{\mu s}|
\approx 10^{-3}\ eV^2, \sin^2 2\theta \sim 10^{-5}$\footnote{
Note that the additional oscillation mode, $\nu_e \to \nu_s$
could increase the `allowed region' in Figure 2 a little, but
not substantially.}), we expect
the $\nu_\mu \to \nu_s$ oscillations to 
generate a $L_{\nu_\mu}$ asymmetry such that $L^{(\mu)} \to 0$. 
This means that we expect that the $\nu_\mu \to \nu_s$
oscillations will eventually fully populate the sterile
neutrino at a temperature around $6-10$ MeV.
We conclude that this scheme does not seem to be
compatible with a BBN bound of $3.6$ neutrinos.
Hence, if experimental data indicate that $\nu_\mu \to \nu_s$
oscillations are required to explain the atmospheric
neutrino data, and if the solar and LSND anomalies have
been correctly interpreted in terms of neutrino
oscillations, then $N_{eff}^{BBN} < 4$
actually suggests the need for more than four neutrinos.
Of course, if light sterile neutrinos exist then
the theoretically most attractive possibility 
is that there are three of them, just like there
are three types of ordinary neutrinos and three types of charged
leptons etc.  Furthermore there are well motivated
extensions of the standard model which have three
light sterile neutrinos. For example, parity 
conserving theories\cite{fv}
(models with similar neutrino phenomenology
but without a mirror sector have also been
proposed in Ref.\cite{model}) necessarily have three light
sterile neutrinos which can naturally 
explain all of the neutrino anomalies
if neutrinos have mass (see the second paper of Ref.\cite{fv} for
a summary). These 
schemes also have the nice property that they can 
explain these neutrino anomalies without assuming any large mixing 
between generations, and they can also have a neutrino mass 
heirarchy.  These last two features are really expected given
the situation with the quarks and the charged leptons.
The implications of these models for early Universe cosmology
is in progress\cite{fv5}.

\acknowledgments{
The author would like to thank Ray Volkas for many
discussions (and previous collaboration) on topics discussed
in this paper and for useful comments on a 
preliminary version of this paper.
Discussions with Nicole Bell, Neil Cornish, Osamu Yasuda
and the participants of the {\it Neutrinos 98 conference}
are also greatfully acknowledged.
R. F. is an Australian Research Fellow.}

\newpage
\vskip 0.5cm
\noindent
{\bf Figure Captions}
\vskip 0.5cm
\noindent
Figure 1. The creation of $|L_{\nu_\tau}|/h$ by $\nu_\tau \to 
\nu_s$ oscillations as a function of temperature, $T/MeV$.
In these examples, $\sin^2 2\theta_0 = 10^{-8}$ and
$\delta m^2/eV^2 = -0.5,\ -50,\ -5000$
for the dashed line, solid line and dotted line respectively.
\vskip 0.5cm
\noindent
Figure 2. The `allowed region' of parameter space in the
$\sin^2 2\theta_0^{\tau s}, \ -\delta m^2_{\tau s}/eV^2$ 
plane, where the $\nu_\tau \to \nu_s$ oscillations create
tau lepton number in such a way as to prevent maximal 
$\nu_\mu \to \nu_s$ oscillations from populating the
sterile neutrino for $T \stackrel{>}{\sim} 0.5\ MeV$.
The bold solid line, upper solid line, lower solid line
are the boundaries assuming $|\delta m^2_{atmos}|/eV^2 = 
10^{-2.5}, \ 10^{-2}, \ 10^{-3}$ respectively.
The dashed-dotted line is the bound
Eq.(\ref{yyy}) which arises by requiring that the
$\nu_\tau \to \nu_s$ oscillations do not populate the
sterile neutrino in the temperature region before the 
exponential growth of $L_{\nu_\tau}$ occurs. 
The shaded region is the favoured region in hot + cold
dark matter models (see the discussion in Section VI).
\vskip 0.5cm
\noindent
Figure 3. Detailed predictions for $N_{eff}^{BBN}$
in the $\sin^2 2\theta_0^{\tau s}, 
-\delta m^2_{\tau s}/eV^2$ plane.
For simplicity, $|\delta m^2_{atmos}| = 10^{-2.5}\ eV^2$ 
is used. In this figure $L_{\nu_e} > 0$ is assumed.
Note that $N_{eff}^{BBN}$ is actually
continuous across the horizontal boundary at
$\delta m^2 = -1000 \ eV^2$.  The transition region is 
roughly $400 \stackrel{<}{\sim} -\delta m^2/eV^2
\stackrel{<}{\sim} 3000$ (the transition region is
not shown for simplicity).

\vskip 0.5cm
\noindent
Figure 4. Same as Figure 3, except that
$L_{\nu_e} < 0$ is assumed.

\newpage
\epsfig{file=f1.eps,width=15cm}
\newpage
\epsfig{file=f2.eps,width=15cm}
\newpage
\epsfig{file=f3.eps,width=15cm}
\newpage
\epsfig{file=f4.eps,width=15cm}

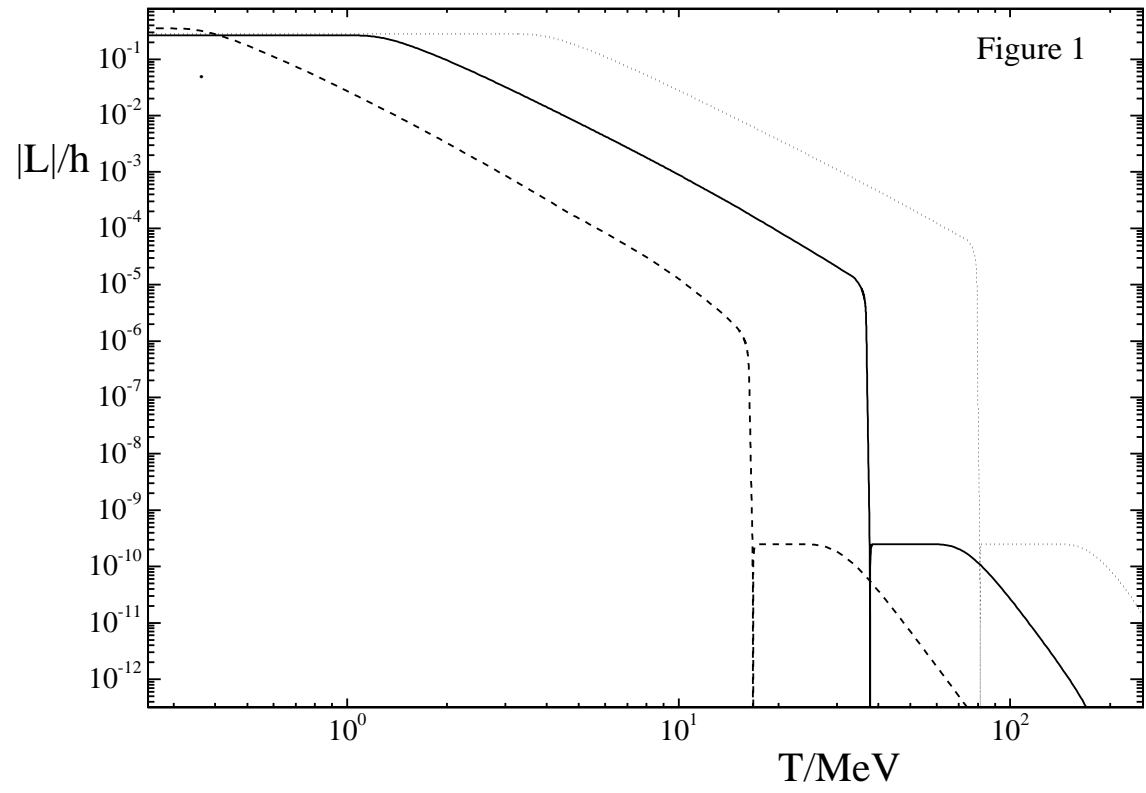
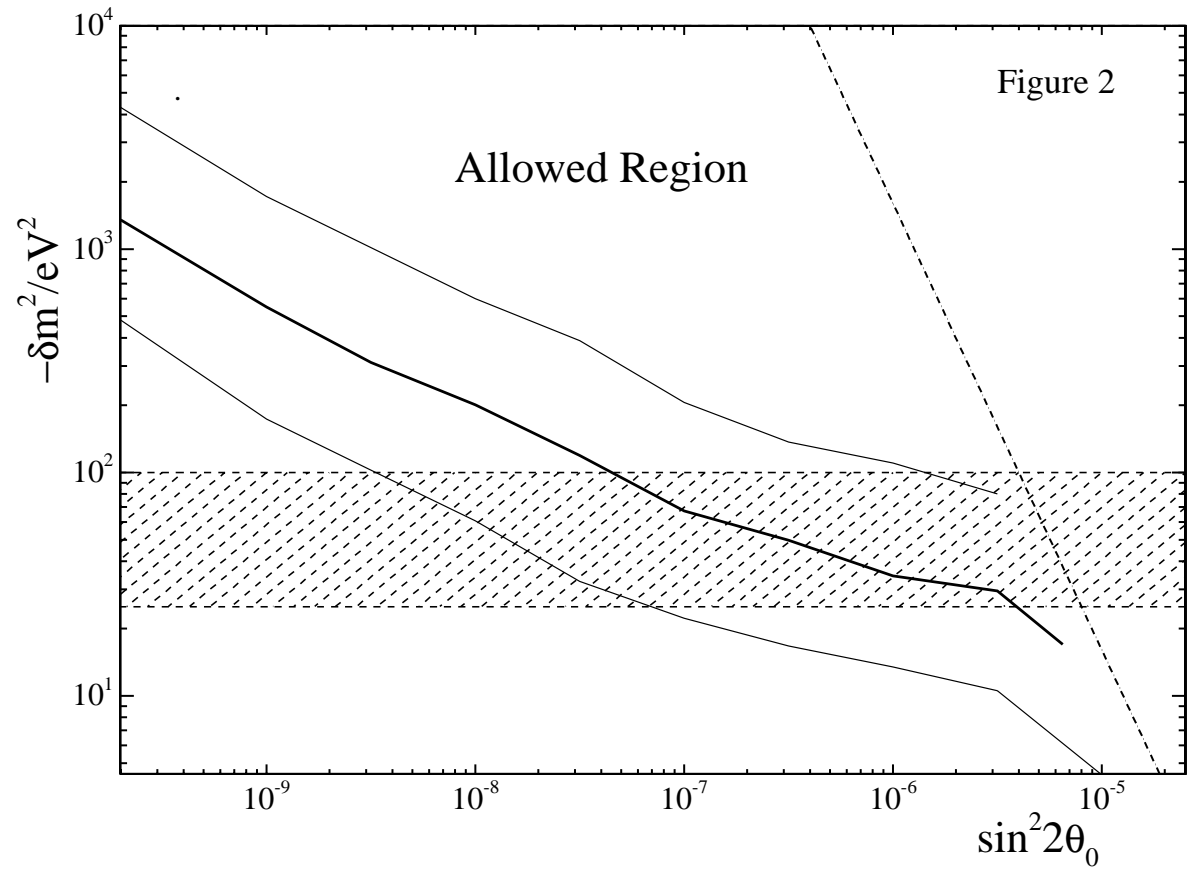
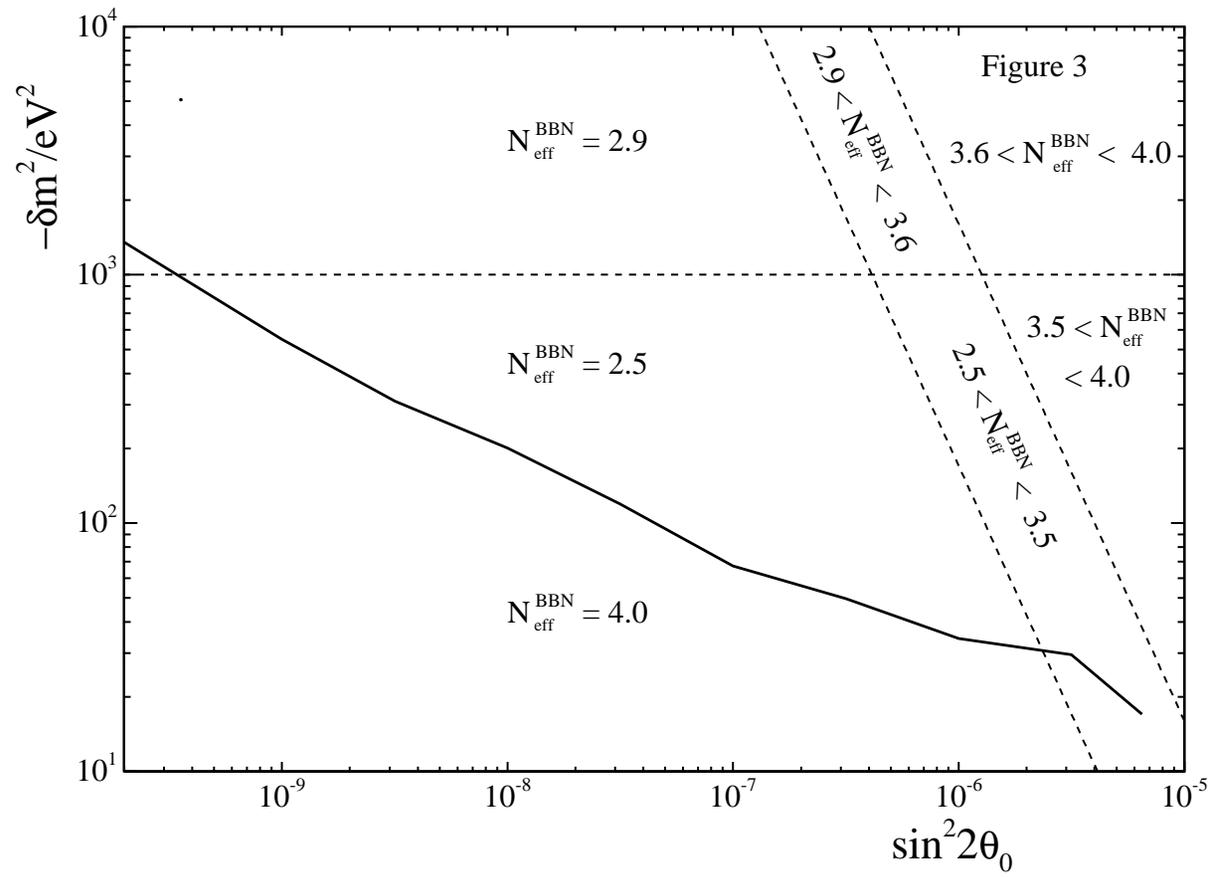
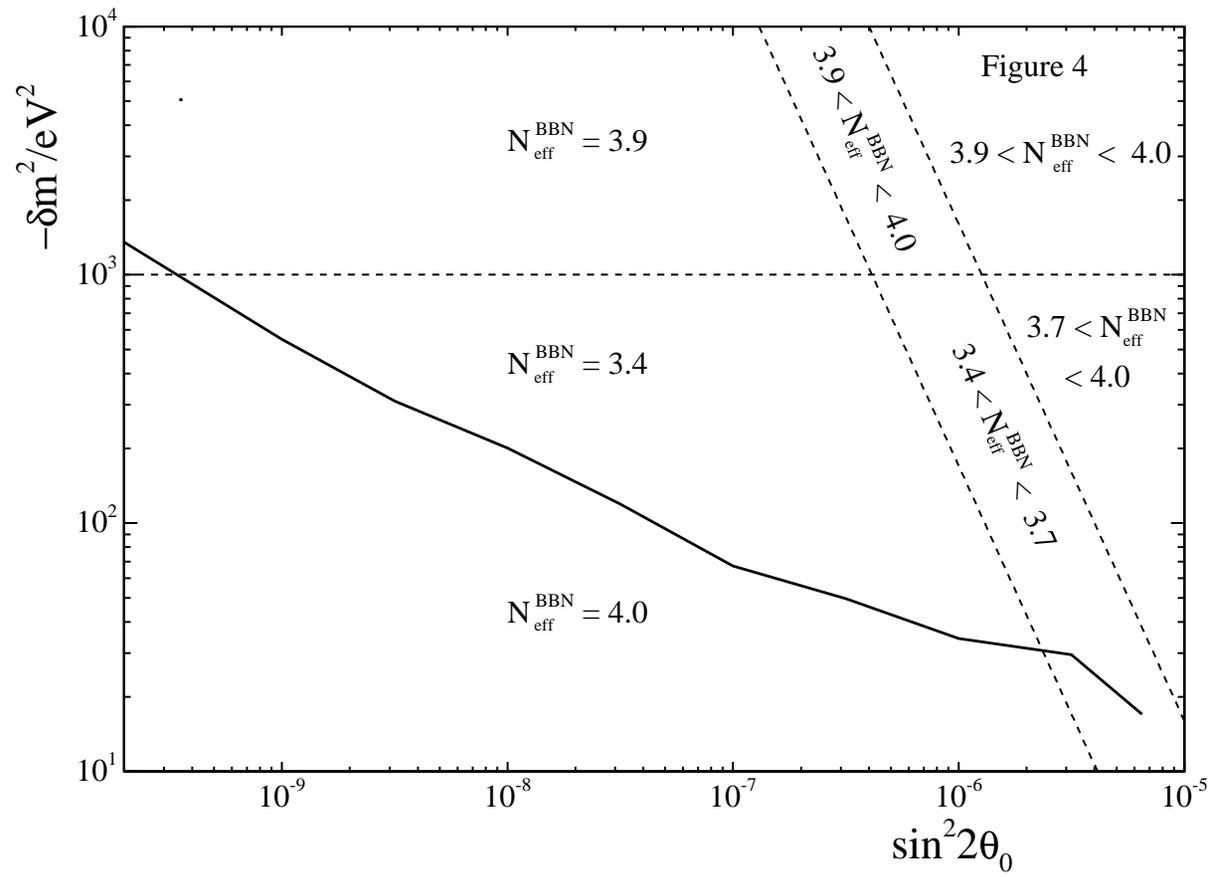
\end{document}